\documentstyle{article}
\begin{document}

\input{psfig}
\title{Symbolic dynamics analysis of the Lorenz equations}

\author{Bai-lin Hao\thanks{On leave from the Institute of Theoretical Physics,
P. O. Box 2735, Beijing 100080, China}\\
Department of Physics, National Central University, Chung-li, Taiwan 32054\\
 Jun-Xian Liu\\Center for Theoretical Physics,
 University of Maribor, Slovenia\\
 and Wei-mou Zheng\\
Institute of Theoretical Physics, P. O. Box 2735, Beijing 100080, China}

\date{\ }
\maketitle

\begin{abstract}
Recent progress of symbolic dynamics of one- and especially two-dimensional
maps has enabled us to construct symbolic dynamics for systems of ordinary
differential equations (ODEs). Numerical study under the guidance of symbolic
dynamics is capable to yield global results on chaotic and periodic regimes
in systems of dissipative ODEs which cannot be obtained neither by purely
analytical means nor by numerical work alone. By constructing symbolic dynamics
of 1D and 2D maps from the Poincar\'e sections all unstable periodic orbits up
to a given length at a fixed parameter set may be located and all stable
periodic orbits up to a given length may be found in a wide parameter range.
This knowledge, in turn, tells much about the nature of the chaotic limits.
Applied to the Lorenz equations, this approach has led to a nomenclature,
i.e., absolute periods and symbolic names, of stable and unstable periodic
orbits for an autonomous system. Symmetry breakings and restorations as well
as coexistence of different regimes are also analyzed by using symbolic
dynamics.
\end{abstract}

\section*{I. INTRODUCTION}

Many interesting nonlinear models in physical sciences and engineering are
given by systems of ODEs. When studying these systems it is desirable to have
a global understanding of the bifurcation and chaos ``spectrum'': the
systematics of periodic orbits, stable as well as unstable ones at fixed and
varying parameters, the type of chaotic attractors which usually occur as
limits of sequences of periodic regimes, etc. However, this is by far not a
simple job to accomplish neither by purely analytical means nor by numerical
work alone. In analytical aspect, just recollect the long-standing problem of
the number of limit cycles in {\em planar} systems of ODEs. As chaotic
behavior may appear only in systems of more than three autonomous ODEs, it
naturally leads to problems much more formidable than counting the number of
limit cycles in planar systems. As numerical study is concerned, one can never
be confident that all stable periodic orbits up to a certain length have been
found in a given parameter range or no short unstable orbits in a chaotic
attractor have been missed at a fixed parameter set, not to mention that it is
extremely difficult to draw global conclusions from numerical data alone.

On the other hand, a properly constructed symbolic dynamics, being a
coarse-grained description, provides a powerful tool to capture global,
topological aspects of the dynamics. This has been convincingly shown in the
development of symbolic dynamics of one-dimensional (1D) maps, see, e.g.,
\cite{mss,ce80,mt88,h89,zh90}. Since it is well known from numerical
observations that chaotic attractors of many higher-dimensional dissipative
systems with one positive Lyapunov exponent reveal 1D-like structure in some
Poincar\'e sections, it has been suggested to associate the systematics of
numerically found periodic orbits in ODEs with symbolic dynamics of 1D maps
\cite{h86}. While this approach has had some success (see, e.g., Chapter~5 of
\cite{h89}), many new questions arose from the case studies. For example,\\
\indent 1. The number of short stable periodic orbits found in ODEs is usually
less than that allowed by the admissibility conditions of the corresponding 1D
symbolic dynamics. Within the 1D framework it is hard to tell whether a missing
period was caused by insufficient numerical search or was forbidden by the
dynamics.

2. In the Poincar\'e sections of ODEs, at a closer examination, the
attractors often reveal two-dimensional features such as layers and folds.
One has to explain the success of 1D description which sometimes even
turns out much better than expected. At the same time, the limitation of 1D
approach has to be analyzed as the Poincar\'e maps are actually two-dimensional.

3. Early efforts were more or less concentrated on stable orbits, while
unstable periods play a fundamental role in organizing chaotic motion.
One has to develop symbolic dynamics for ODEs which would be capable to
treat stable and unstable periodic orbits alike, to indicate the
structure of some, though not all, chaotic orbits at a given parameter set.

The elucidation of these problems has to await a significant progress of
symbolic dynamics of 2D maps. Now the time is ripe for an in depth symbolic
dynamics analysis of a few typical ODEs. This kind of analysis has been
carried out on several non-autonomous systems \cite{xzh95,lzh96,lwz96}, where
the stroboscopic sampling method \cite{hz82} greatly simplifies the
calculation of Poincar\'e maps. In this paper we consider an autonomous system,
namely, the Lorenz model, in which one of the first chaotic attractor was
discovered \cite{l63}.

The Lorenz model consists of three equations
\begin{equation}
{\dot x}=\sigma(y-x), \quad {\dot y}=rx-y-xy, \quad {\dot z}=xy-bx.
\label{lorenz}
\end{equation}
It is known that several models of hydrodynamical, mechanical, dynamo and laser
problems may be reduced to this set of ODEs. The system (\ref{lorenz}) contains
three parameters $r$, $\sigma$ and $b$, representing respectively the Rayleigh
number, the Prandtl number and a geometric ratio. We will study the system in
a wide $r$-range at fixed $\sigma=10$ and $b=8/3$.

We put together a few known facts on Eq.~\ref{lorenz} to fix the notations.
For detailed derivations one may refer to the book \cite{s82} by C. Sparrow.
For $0 < r < 1$ the origin $(0, 0, 0)$ is a globally stable fixed point.
It loses stability at $r=1$. A 1D unstable manifold and a 2D stable manifold
${\cal W}^s$ come out from the unstable origin. The intersection of the
2D ${\cal W}^s$ with the Poincar\'e section will determine a demarcation line
in the partition of the 2D phase plane of the Poincar\'e map. For $r > 1$ 
there appears a pair of fixed points
\[ C_\pm =(\pm\sqrt{b(r-1)}, \pm\sqrt{b(r-1)}, r-1). \]
These two fixed points remain stable until $r$ reaches 24.74. Although their
eigenvalues undergo some qualitative changes at $r=1.345617$ and a
strange invariant set (not an attractor yet) comes into life at $r=13.926$,
here we are not interested in all this. It is at $r=24.74$ a sub-critical Hopf
bifurcation takes place and chaotic regimes commence. Our $r$-range extends
from 28 to very big values, e.g., 10000, as nothing qualitatively new appears
at, say, $r>350$.

Before undertaking the symbolic dynamics analysis we summarize briefly what
has been done on the Lorenz system from the viewpoint of symbolic dynamics.
Guckenheimer and Williams introduced the geometric Lorenz model \cite{gw79}
for the vicinity of $r=28$ which leads to symbolic dynamics on two letters,
proving the existence of chaos in the geometric model. However, as Smale
\cite{s91} pointed out it remains an unsolved problem as whether the geometric
Lorenz model means to the real Lorenz system. Though not using symbolic
dynamics at all, the paper by Tomita and Tsuda \cite{tt80} studying the Lorenz
equations at a different set of parameters $\sigma=16$ and $b=4$ is worth
mentioning. They noticed that the quasi-1D chaotic attractor in the
$z=r-1$ Poincar\'e section outlined by the upward intersections of the
trajectories may be directly parameterized by the $x$ coordinates. A 1D map
was devised in \cite{tt80} to numerically mimic the global bifurcation
structure of the Lorenz model. C. Sparrow \cite{s82} used two symbols $x$ and
$y$ to encode orbits without explicitly constructing symbolic dynamics.
In Appendix~J of \cite{s82} Sparrow described a family of 1D maps as ``an
obvious choice if we wish to try and model the behavior of the Lorenz equations
in the parameter range $\sigma=10$, $b=8/3$ and $r>24.06$''. In what follows we
will call this family the {\em Lorenz-Sparrow map}. Refs.~\cite{tt80}
and~\cite{s82} have been instrumental for the present study. In fact, the 1D
maps to be obtained from the 2D upward Poincar\'e maps of the Lorenz equations
after some manipulations belong precisely to the family suggested by Sparrow.
In \cite{dh88} the systematics of stable periodic orbits in the Lorenz equations
was compared with that of a 1D anti-symmetric cubic map. The choice of an
anti-symmetric map was dictated by the invariance of the Lorenz equations under
the discrete transformation
\begin{equation}
\label{invers}
x\rightarrow -x,\quad y\rightarrow -y,\quad {\rm and}\quad z\rightarrow z.
\end{equation}
Indeed, most of the periods known to \cite{dh88} are ordered in a ``cubic'' way.
However, many short periods present in the 1D map have not been found in the
Lorenz equations. It was realized in \cite{fh96} that a cubic map with a
discontinuity in the center may better reflect the ODEs and many of the
missing periods are excluded by the 2D nature of the Poincar\'e map. Instead
of devising model maps for comparison one should generate all related 1D or 2D
maps directly from the Lorenz equations and construct the corresponding
symbolic dynamics. This makes the main body of the present paper.

\begin{figure}[t]
\centerline{\psfig{figure=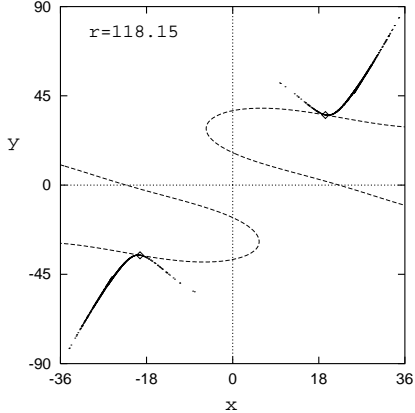,width=6cm,height=6cm}}
\caption{An upward Poincar\'e section at $r=118.15$. The dashed curve is one
of the forward contracting foliations and the diamond a tangent point between
the FCF and BCF.}
\label{hlzfig1}
\end{figure}

For physicists symbolic dynamics is nothing but a coarse-grained description
of the dynamics. The success of symbolic dynamics depends on how the
coarse-graining is performed, i.e., on the partition of the phase space.
From a practical point of view we can put forward the following requirements
for a good partition. 1. It should assign a {\em unique} name to each
unstable periodic orbit in the system; 2. An ordering rule of all symbolic
sequences should be defined; 3. Admissibility conditions as whether a given
symbolic sequence is allowed by the dynamics should be formulated; 4. Based on
the admissibility conditions and ordering rule one should be able to generate
and locate all periodic orbits, stable and unstable, up to a given length.
Symbolic dynamics of 1D maps has been well understood
\cite{mss,ce80,mt88,h89,zh90}. Symbolic dynamics of 2D maps has been studies
in \cite{pv87,cgp88,gkm89,bw90,agsp90,aip91,z9192,z92,zzg92,zz93}. We will
explain the main idea and technique in the context of the Lorenz equations.

A few words on the research strategy may be in order. We will first
calculate the Poincar\'e maps in suitably chosen sections. If necessary some
forward contracting foliations (FCFs, to be explained later) are superimposed
on the Poincar\'e map, the attractor being part of the backward contracting
foliations (BCFs). Then a one-parameter parameterization is introduced for the
quasi-1D attractor. For our choice of the Poincar\'e sections the
parameterization is simply realized by the $x$ coordinates of the points.
In terms of these $\{ x_i\}$ a first return map $x_n \mapsto x_{n+1}$ is
constructed. Using the specific property of first return maps that the set
$\{x_i\}$ remains the same before and after the mapping, some parts of
$\{x_i\}$ may be safely shifted and swapped to yield a new map $x_n^\prime
\mapsto x_{n+1}^\prime$, which precisely belongs to the family of Lorenz-Sparrow
map. In so doing, all 2D features (layers, folds, etc.) are kept. However, one
can always start from the symbolic dynamics of the 1D Lorenz-Sparrow map to
generate a list of allowed periods and then check them against the admissibility
conditions of the 2D symbolic dynamics. Using the ordering of symbolic sequences
all allowed periods may be located easily.  What said applies to unstable
periodic orbits at fixed parameter set. The same method can be adapted to treat
stable periods either by superimposing the orbital points on a near-by chaotic
attractor or by keeping a sufficient number of transient points.

\begin{figure}[t]
\centerline{\psfig{figure=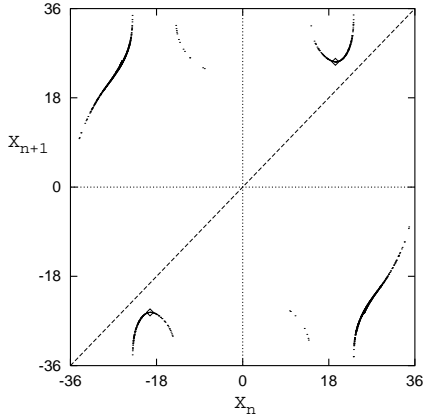,width=6cm,height=6cm}}
\caption{The first return map $x_n \mapsto x_{n+1}$ constructed from
 Fig.~\ref{hlzfig1} by using the $x$ coordinates.}
\label{hlzfig2}
\end{figure}

\section*{II. CONSTRUCTION OF POINCAR\'E AND\protect\\
 RETURN MAPS}

The Poincar\'e map in the $z=r-1$ plane captures most of the interesting
dynamics as it contains both fixed points $C_\pm$. The $z$-axis is contained
in the stable manifold ${\cal W}^s$ of the origin $(0, 0, 0)$. All orbits
reaching the $z$-axis will be attracted to the origin , thus most of
the homoclinic behavior may be tracked in this plane. In principle, either
downward or upward intersections of trajectories with the $z=r-1$ plane
may be used to generate the Poincar\'e map. However, upward intersections
with $dz/dt>0$ have the practical merit to yield 1D-like objects which may
be parameterized by simply using the $x$ coordinates.

\begin{figure}[t]
\centerline{\psfig{figure=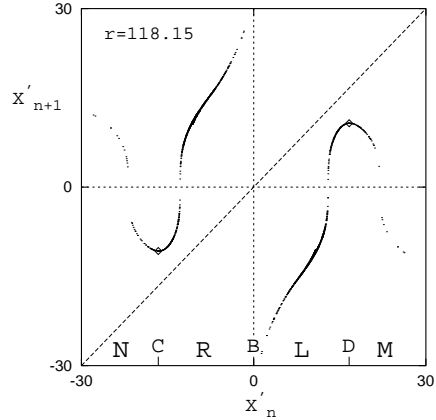,width=6cm,height=6cm}}
\caption{The swapped return map $x_n^\prime\mapsto x_{n+1}^\prime$ constructed
from Fig.~\ref{hlzfig2}. The gaps may be further squeezed, see text.}
\label{hlzfig3}
\end{figure}

Fig.~\ref{hlzfig1} shows a Poincar\'e section at $r=118.15$. The dashed curves
and diamonds represent one of the FCFs and its tangent points with the BCF.
These will be used later in Sec.~V. The 1D-like structure of the attractor is
apparent. Only the thickening in some part of the attractor hints on 2D
structures. Ignoring the thickening for the time being, the 1D attractor may be
parameterized by the $x$ coordinates only. Collecting successive $x_i$, we
construct a first return map $x_n \mapsto x_{n+1}$ as shown in
Fig.~\ref{hlzfig2}. It consists of four symmetrically located pieces with gaps
on the mapping interval. For a first return map a gap belonging to both
$\{x_n\}$ and $\{x_{n+1}\}$ plays no role in the dynamics. If necessary, we can
use this specificity of return maps to squeeze some gaps in $x$. Furthermore,
we can interchange the left subinterval with the right one by defining, e.g.,
\begin{equation}
\label{swap}
x^\prime=x - 36\quad {\rm for} \quad x>0; \qquad x^\prime=x+36\quad
 {\rm for} \quad x<0.
\end{equation}
The precise value of the numerical constant is not essential; it may be
estimated from the upper bound of $\{|x_i|\}$ and is so chosen as to make the
final figure look nicer. The swapped first return map, as we call it, is shown
in Fig.~\ref{hlzfig3}. The corresponding tangent points between FCF and BCF
(the diamonds) are also drawn on these return maps for later use.

\begin{figure}[p]
\centerline{\psfig{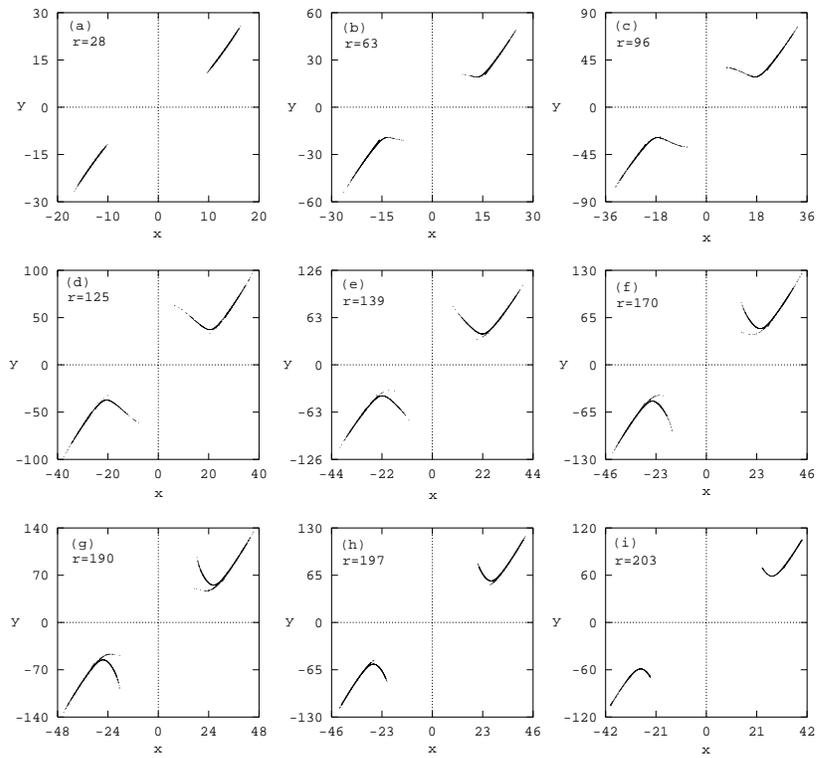}}
\caption{Upward Poincar\'e maps at 9 different $r$-values.}
\label{hlzfig4}
\end{figure}

\begin{figure}[p]
\centerline{\psfig{figure=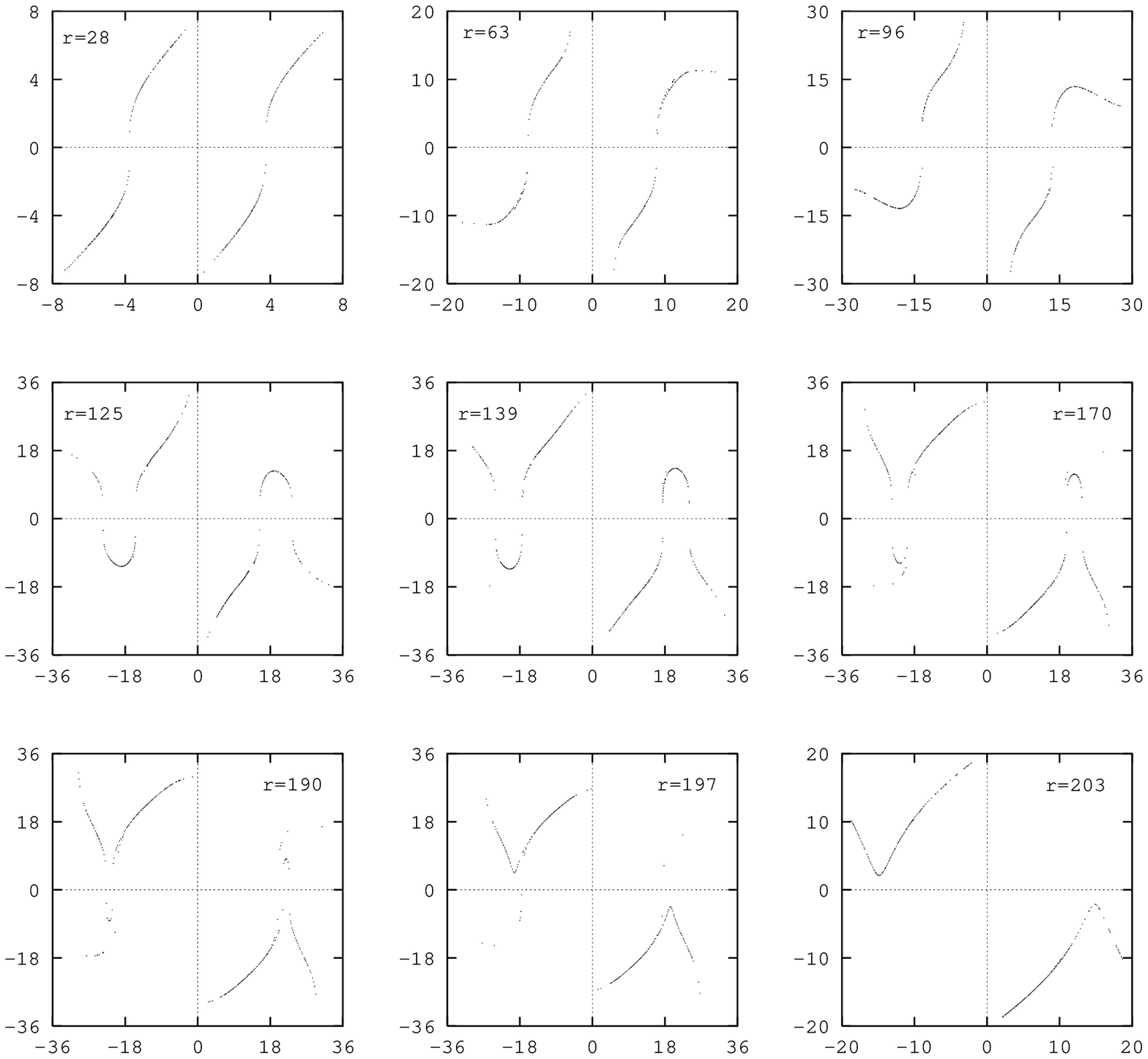,width=12cm,height=11.14cm}}
\caption{Swapped first return maps obtained from the Poincar\'e maps shown
 in Fig.~\ref{hlzfig4}.}
\label{hlzfig5}
\end{figure}

It is crucial that the parameterization and swapping do keep the 2D features
present in the Poincar\'e map. This is important when it comes to take into
account the 2D nature of the Poincar\'e maps.

In Fig.~\ref{hlzfig4} Poincar\'e maps at 9 different values from $r=28$ to 203
are shown. The corresponding swapped return maps are shown in
Fig.~\ref{hlzfig5}. Generally speaking, as~$r$ varies from small to greater
values, these maps undergo transitions from 1D-like to 2D-like, and then to
1D-like again. Even in the 2D-like range the 1D backbones still dominate.
This partly explains our early success \cite{dh88,fh96} in applying purely 1D
symbolic dynamics to the Lorenz model. We will learn how to judge this success
later on. Some qualitative changes at varying~$r$ will be discussed in
Sec.~III. We note also that the return map at $r=28$ complies with what
follows from the geometric Lorenz model. The symbolic dynamics of this
Lorenz-like map has been completely constructed \cite{z90}.

\section*{III. SYMBOLIC DYNAMICS OF THE\protect\\
 1D LORENZ-SPARROW MAP}

All the return maps shown in Fig.~\ref{hlzfig5} fit into the family of
Lorenz-Sparrow map. Therefore, we take a general map from the family and
construct the symbolic dynamics. There is no need to have analytical expression
for the map. Suffice it to define a map by the shape shown in
Fig.~\ref{hlzfig6}. This map has four monotone branches, defined on four
subintervals labeled by the letters $M$, $L$, $N$, and $R$, respectively.
We will also use these same letters to denote the monotone branches themselves,
although we do not have an expression for the mapping function $f(x)$. Among
these branches $R$ and $L$ are increasing; we say $R$ and $L$ have an even or
$+$ {\em parity}. The decreasing branches $M$ and $N$ have odd or $-$ parity.
Between the monotone branches there are ``turning points'' (``critical points'')
$D$ and $C$ as well as ``breaking point'' $B$, where a discontinuity is present.
Any numerical trajectory $x_1x_2\cdots x_i\cdots$ in this map corresponds to a
symbolic sequence
\[ \Sigma=\sigma_1\sigma_2\cdots\sigma_i\cdots, \]
where $\sigma_i\in\{M, L, N, R, C, D, B\}$, depending on where the point $x_i$
falls in.

\subsection*{Ordering and admissibility of symbolic sequences}

All symbolic sequences made of these letters may be ordered in the following
way. First, there is a natural order
\begin{equation}
\label{order1}
N < C < R < B < L < D < M.
\end{equation}
Next, if two symbolic sequences $\Sigma_1$ and $\Sigma_2$ have a common
leading string $\Sigma^\ast$, i.e.,
\[ \Sigma_1 = \Sigma^\ast\sigma\cdots,\quad \Sigma_2 = \Sigma^\ast\tau\cdots,
\]
where $\sigma\neq \tau$. Since $\sigma$ and $\tau$ are different, they must
have ordered according to (\ref{order1}). The {\em ordering rule} is:
if $\Sigma$ is even, i.e., it contains an even number of $N$ and $M$, the
order of $\Sigma_1$ and $\Sigma_2$ is given by that of $\sigma$ and $\tau$;
if $\Sigma^\ast$ is odd, the order is the opposite to that of $\sigma$
and $\tau$. The ordering rule may be put in the following form:
\begin{eqnarray}
\label{forder1}
EN\cdots &<& EC\cdots < ER\cdots < EB\cdots\nonumber\\
 {}      &<& EL\cdots < ED\cdots < EM\cdots,\nonumber\\
ON\cdots &>& OC\cdots > OR\cdots > OB\cdots\nonumber\\
 {}      &>& OL\cdots > OD\cdots > OM\cdots,\nonumber\\
\end{eqnarray}
where $E$ ($O$) represents a finite string of $M$, $L$, $N$, and $R$
containing an {\em even} ({\em odd}) number of letters $M$ and $N$. We call
$E$ and $O$ even and odd string, respectively.

\begin{figure}
\centerline{\psfig{figure=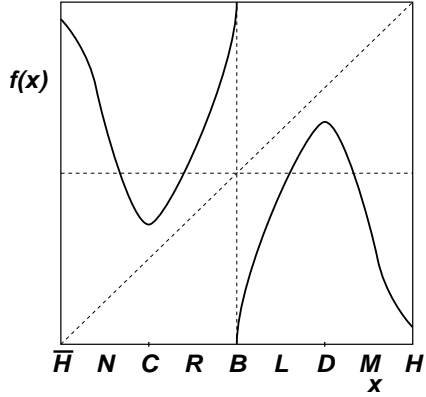,width=6cm,height=6cm}}
\caption{A generic Lorenz-Sparrow map. The symbols $M$, $L$, $R$, and $N$
label the monotone branches as well as the subintervals. $C$, $D$ and $B$
are turning or breaking points. For $H$ and $\overline{H}$ see text.}
\label{hlzfig6}
\end{figure}

In order to incorporate the discrete symmetry, we define a transformation
${\cal T}$ of symbols:
\begin{equation}
\label{trans}
{\cal T} =\{ M \leftrightarrow N, L\leftrightarrow R, C\leftrightarrow D\},
\end{equation}
keeping $B$ unchanged. Sometimes we distinguish the left and right limit of
$B$, then we add $B_-\leftrightarrow B_+$. We often denote ${\cal T}\Sigma$
by $\overline{\Sigma}$ and say $\Sigma$ and $\overline{\Sigma}$ are mirror
images to each other.

Symbolic sequences that start from the next iterate of the turning or
breaking points play a key role in symbolic dynamics. They are called
{\em kneading sequences} \cite{mt88}. Naming a symbolic sequence by the
initial number which corresponds to its first symbol, we have two kneading
sequences from the turning points:
\[ K=f(C),\qquad \overline{K}=f(D).\]
Being mirror images to each other, we take $K$ as the independent one.

For first return maps the rightmost point in $\{x_i\}$ equals the highest
point after the mapping. Therefore, $f(B_-)=H$ and $f(B_+)=\overline{H}$, see
Fig.~\ref{hlzfig6}. We take $H$ as another kneading sequence. Note that $B_-$
and $B_+$ are not necessarily the left and right limit of the breaking point;
a finite gap may exist in between. This is associated with the flexibility of
choosing the shift constant, e.g., the number 36 in (\ref{swap}).
Since a kneading sequence starts from the first iterate of a turning or
breaking point, we have
\begin{eqnarray}
\label{knead}
C_-&=&NK,\quad C+=RK,\quad B_-=RH,\nonumber\\
B_+&=&L\overline{H},\quad D_-=L\overline{K},\quad D_+=M\overline{K}\nonumber.\\
\end{eqnarray}

A 1D map with multiple critical points is best parameterized by its kneading
sequences. The dynamical behavior of the Lorenz-Sparrow map is entirely
determined by a {\em kneading pair} $(K, H)$. Given a kneading pair $(K, H)$,
not all symbolic sequences are allowed in the dynamics. In order to formulate
the admissibility conditions we need a new notion. Take a symbolic
sequence $\Sigma$ and inspect its symbols one by one. Whenever a letter $M$
is encountered, we collect the subsequent sequence that follows this $M$. The
set of all such sequences is denoted by ${\cal M}(\Sigma)$ and is called a
$M$-shift set of $\Sigma$. Similarly, we define ${\cal L}(\Sigma)$,
${\cal R}(\Sigma)$ and ${\cal N}(\Sigma)$.

The {\em admissibility conditions}, based on the ordering rule (\ref{forder1}),
follow from (\ref{knead}): 
\begin{eqnarray}
\label{admis}
\overline{H} &\leq& N{\cal N}(\Sigma)\leq NK,\quad
K\leq {\cal R}(\Sigma)\leq H,\nonumber\\ 
\overline{H} &\leq& {\cal L}(\Sigma)\leq \overline{K},\quad
M\overline{K}\leq M{\cal M}(\Sigma)\leq H\nonumber.\\
\end{eqnarray}
Here in the two middle relations we have canceled the leading $R$ or $L$.

The twofold meaning of the admissibility conditions should be emphasized.
On one hand, for a given kneading pair these conditions
select those symbolic sequences which may occur in the dynamics. On the other
hand, a kneading pair $(K, H)$, being symbolic sequences themselves, must also
satisfy conditions (\ref{admis}) with $\Sigma$ replaced by $K$ and $H$. Such
$(K, H)$ is called a {\em compatible} kneading pair. The first meaning concerns
admissible sequences in the phase space at a fixed parameter set while the
second deals with compatible kneading pairs in the parameter space. In
accordance with these two aspects there are two pieces of work to be done.
First, generate all compatible kneading pairs up to a given length. This is
treated in Appendix~A. Second, generate all admissible symbolic sequences up
to a certain length for a given kneading pair $(K, H)$. The procedure is
described in Appendix~B. 

\subsection*{Metric representation of symbolic sequences}

\begin{figure}[t]
\centerline{\psfig{figure=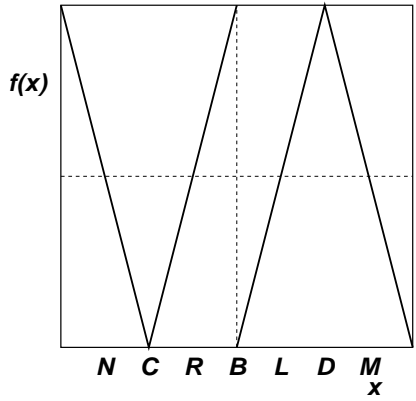,width=6cm,height=6cm}}
\caption{A piecewise linear map used to introduce metric representation
for the Lorenz-Sparrow map.}
\label{hlzfig7}
\end{figure}

It is convenient to introduce a metric representation of symbolic sequences
by associating a real number $0\leq\alpha(\Sigma)\leq 1$ to each sequence
$\Sigma$. To do so let us look at the piecewise linear map shown in
Fig.~\ref{hlzfig7}. It is an analog of the surjective tent map in the sense
that all symbolic sequences made of the four letters $M$, $L$, $R$, and $N$
are allowed. It is obvious that the maximal sequence is $(MN)^\infty$
while the minimal one being $(NM)^\infty$. For this map one may further write
\begin{eqnarray*}
C_-=N(NM)^\infty, C+=R(NM)^\infty, B_-=R(MN)^\infty,\\
B_+=L(NM)^\infty, D_-=L(MN)^\infty, D_+=M(MN)^\infty.\\
\end{eqnarray*}

To introduce the metric representation we first use $\epsilon=1$ to mark the
even parity of $L$ and $R$, and $\epsilon=-1$ to mark the odd parity of $M$
and $N$. Next, the number $\alpha(\Sigma)$ is defined for a sequence 
$\Sigma=s_1s_2 \cdots s_i\cdots$ as
\begin{equation}
\label{fmetric}
\alpha =\sum_{i=1}^\infty \mu_i 4^{-i},
\end{equation}
where
\[ \mu_i = \cases{0\cr 1\cr 2\cr 3\cr} \quad {\rm for }\quad
s_i=  \cases{N\cr R\cr L\cr M\cr} \quad {\rm if} \quad
\epsilon_1\epsilon_2\cdots \epsilon_{i-1}=1, \]
or,
\[ \mu_i= \cases{3\cr 2\cr 1\cr 0\cr} \quad {\rm for }\quad
s_i=  \cases{N\cr R\cr L\cr M\cr} \quad {\rm if} \quad
\epsilon_1\epsilon_2\cdots \epsilon_{i-1}=-1. \]
It is easy to check that
\begin{eqnarray*}
\alpha((NM)^\infty)&=&0,\quad \alpha(C_\pm)=1/4,\quad \alpha(B_\pm)=1/2,\\
\alpha(D_\pm)&=&3/4,\quad \alpha((MN)^\infty)=1.\\
\end{eqnarray*}
The following relations hold for any symbolic sequence $\Sigma$:
\begin{eqnarray}
\label{alpha}
\alpha(\overline{\Sigma})=1-\alpha(\Sigma),\quad \alpha(L\Sigma)
&=&(2+\alpha(\Sigma))/4,\nonumber\\
\alpha(R\Sigma)&=&(1+\alpha(\Sigma))/4\nonumber.\\
\end{eqnarray}
One may also formulate the admissibility conditions in terms of the metric
representations.

\subsection*{One-parameter limits of the Lorenz-Sparrow map}

The family of the Lorenz-Sparrow map includes some limiting cases.\\
\indent{1}. The $N$ branch may disappear, and the minimal point of the $R$
branch moves to the left end of the interval. This may be described as
\begin{equation}
\label{onep1}
C=\overline{H}=RK,\quad {\rm or}\quad H=L\overline{K}.
\end{equation}
It defines the only kneading sequence $K$ from the next iterate of $C$.

2. The minimum at $C$ may rise above the horizontal axis, as it is evident in
Fig.~\ref{hlzfig5} at $r=203$. The second iterate of either the left or right
subinterval then retains in the same subinterval. Consequently, the two
kneading sequences are no longer independent and they are bound by the relation
\begin{equation}
\label{onep2}
K=L\overline{H},\quad {\rm or}\quad \overline{K}=RH.
\end{equation}

Both one-parameter limits appear in the Lorenz equations as we shall see in
the next section.

\section*{IV. 1D SYMBOLIC DYNAMICS OF THE\protect\\
 LORENZ EQUATIONS}

Now we are well prepared to carry out a 1D symbolic dynamics analysis of the
Lorenz equations using the swapped return maps shown in Fig.~\ref{hlzfig5}.
We take $r=118.15$ as an working example. The rightmost point in $\{x_i\}$ and
the minimum at $C$ determine the two kneading sequences:
\begin{eqnarray*}
H & = & MRLNRLRLRLRLNRL\cdots,\\
K & = & RLRLRLRRLRLRLRR\cdots.\\
\end{eqnarray*}
Indeed, they satisfy (\ref{admis}) and form a compatible kneading pair. Using
the propositions formulated in Appendix~B, all admissible periodic sequences up
to period~6 are generated. They are $LC$, $LNLR$, $LNLRLC$, $RMRLR$, $RMLNLC$,
$RMLN$, $RLLC$, $RLLNLC$, $RLRMLC$, and $RLRLLC$. Here the letter $C$ is used to
denote both $N$ and $R$. Therefore, there are altogether 17 unstable periodic
orbits with period equal or less than~6. Relying on the ordering of symbolic
sequences and using a bisection method, these unstable periodic orbits may be
quickly located in the phase plane.

It should be emphasized that we are dealing with unstable periodic orbits at
a fixed parameter set. There is no such thing as superstable periodic sequence
or periodic window which would appear when one considers kneading sequences
with varying parameters. Consequently, the existence of $LN$ and $LR$ does not
necessarily imply the existence of $LC$.

\begin{figure}[p]
\centerline{\psfig{figure=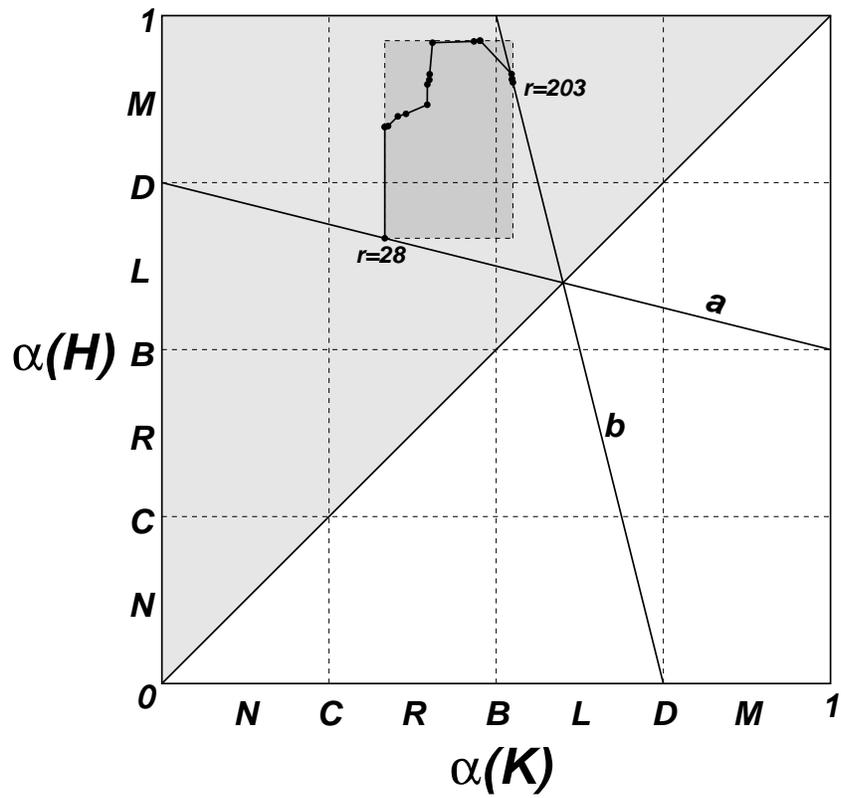,width=12cm,height=11.33cm}}
\caption{The $\alpha(H)$ versus $\alpha(K)$ plane shows kneading pairs (solid
circles) corresponding to the Lorenz equations from $r=28$ to 203. Only the
upper left triangular region is accessible for compatible pairs in the
Loren-Sparrow map. The two straight lines {\bf a} and {\bf b} represent the two
one-parameter limits of the Lorenz-Sparrow map.}
\label{hlzfig8}
\end{figure}

Similar analysis may be carried out for other $r$. In Table~\ref{table1} we
collect some kneading sequences at different $r$-values. Their corresponding
metric representations are also included. We first note that they do satisfy
the admissibility conditions (\ref{admis}), i.e., $K$ and $H$ at each~$r$ make
a compatible kneading pair. An instructive way of presenting the data consists
in drawing the plane of metric representation for both $\alpha(K)$ and
$\alpha(H)$, see Fig.~\ref{hlzfig8}. The compatibility conditions require, in
particular, $K\leq H$, therefore only the upper left triangular region is
accessible.

\begin{table}[p]
\caption{Kneading pairs $(K, H)$ and their metric representations at
 different $r$-values.}
\label{table1}
\begin{center}
\begin{scriptsize}
\begin{tabular}{clclc}
\hline\hline
\multicolumn{1}{c}{$r$} &\multicolumn{1}{c}{$K$} & $\alpha(K)$ &
\multicolumn{1}{c}{$H$} & $\alpha(H)$ \\
\hline 
203.0 & LNLRLRLRLRLRLRL & 0.525000 & MRLRLRLRLRLRLRL & 0.900000 \\
201.0 & LNLRLRLRMRLRLRM & 0.524995 & MRLRLRLNLRLRLNL & 0.900018 \\
199.04 & LNLRLRMRLRLNLRL & 0.524927 & MRLRLNLRLRMRLRL & 0.900293 \\
198.50 & LNLRMRMRLRMRMRL & 0.523901 & MRLNLNLRLNLNLRL & 0.904396 \\
197.70 & LNLRMRLRLRLNLRL & 0.523828 & MRLNLRLRLRMRLRL & 0.904688 \\
197.65 & LNLRMRLRLNLRLRM & 0.523827 & MRLNLRLRMRLRLNL & 0.904692 \\
197.58 & LNLRMRRMRMRLRMR & 0.523796 & MRRMRMRLNLNLRLR & 0.908110 \\
196.20 & LNLLNLRLNLRLRLN & 0.523042 & MRRLRLRLRLRLNLR & 0.912500 \\
191.0 & RMRLNLNLLRLRLRM & 0.476096 & MNRLRLRMRLRMLNL & 0.962518 \\
181.8 & RMRLLNLRRMRLLNL & 0.475410 & MNRLRLRLRLRLRLN & 0.962500 \\
166.2 & RMLNRMLNRMLNRML & 0.466667 & MNRLNRMRLRLLNRM & 0.961450 \\
139.4 & RLRMRLRRLLNLRRL & 0.404699 & MNRRLLNRMLRRLLR & 0.959505 \\
136.5 & RLRMLRRLRMRLLRL & 0.403954 & MNRRLLNLRRLRMLR & 0.959498 \\
125.0 & RLRLLNLRLRRLRLL & 0.400488 & MRRLRMLRLRRLRLL & 0.912256 \\
120.0 & RLRLRLRLRLRLRML & 0.400000 & MRRMLRLRLRLRLRR & 0.908594 \\
118.15 & RLRLRLRRLRLRLRR & 0.399988 & MRLNRLRLRLRLNRL & 0.903906 \\
117.7 & RLRLRLNRLLRLRLR & 0.399938 & MRLRRLRLRLNRLLR & 0.900781 \\
114.02 & RLRLRRLRLRRLRLR & 0.399804 & MRLRLNRLRLRRLRL & 0.900244 \\
107.7 & RLRRLLRRLRRLLRR & 0.397058 & MRLLRRLLRLLNRLR & 0.897058 \\
104.2 & RLRRLRLRRLRRLRL & 0.396872 & MRLLRLLRLRRLLNR & 0.896826 \\
101.5 & RLRRLRRLRRLRLNR & 0.396825 & MLNRMLRLLRLLRLL & 0.866598 \\
99.0 & RLNRLRRLRRLRRLR & 0.384425 & MLNRLRRLNRLRRLN & 0.865572\\
93.4 & RRMLLNRRMLLNRLL & 0.365079 & MLRRLLRRRMLRLLR & 0.852944 \\
83.5 & RRLLRRRLLRRRLLR & 0.352884 & MLRLLRLRRRLLLRR & 0.849233 \\
71.7 & RRLRRRLRRRLRRLR & 0.349020 & MLLRLRRLLRRLLLR & 0.837546\\
69.9 & RRRMLLLNRRRMLLR & 0.341176 & MLLRLRLLLRLLRLR & 0.837485 \\
69.65 & RRRLLLLNRRRMLLR & 0.338511 & MLLRLLNRRLRRLLR & 0.837363 \\
65.0 & RRRLLRRRRLLRLLR & 0.338217 & MLLLRRRLRRRLLRL & 0.834620 \\
62.2 & RRRLRLRRRLRRRRM & 0.337485 & MLLLRRLLLLRLRLL & 0.834554 \\
59.4 & RRRLRRRRLRRRLLL & 0.337243 & MLLLRLLRLRRLRLL & 0.834326 \\
55.9 & RRRRLLRRRRLRLLR & 0.334554 & MLLLLRRLRRLLLLR & 0.833643 \\
52.6 & RRRRLRRRRRLLRRR & 0.334310 & MLLLLRLLLLRRRRR & 0.833578 \\
50.5 & RRRRRLLRRRLRLRL & 0.333639 & MLLLLLRRLLRRLRR & 0.833410 \\
48.3 & RRRRRLRRRRRLLLL & 0.333578 & MLLLLLRLLLLRLLR & 0.833394 \\
48.05 & RRRRRRLLLLLRLLL & 0.333415 & MLLLLLRLLLLLLLR & 0.833394 \\
46.0 & RRRRRRLRLRRRRLL & 0.333398 & MLLLLLLRLRLLLLN & 0.833350 \\
44.0 & RRRRRRRLRLLLLLL & 0.333350 & MLLLLLLLRLRRRRR & 0.833337 \\
36.0 & RRRRRRRRRRRRRLR & 0.333333 & MLLLLLLLLLLLLLR & 0.833333 \\
28.0 & RRRRRRRRRRRRRRR & 0.333333 & LLLLLLLLLLLLLLL & 0.666667 \\
\hline\hline
\end{tabular}
\end{scriptsize}
\end{center}
\end{table}

As we have indicated at the end of the last section, the Lorenz-Sparrow map
has two one-parameter limits. The first limit (\ref{onep1}) takes place
somewhere at $r<36$, maybe around $r < 30.1$, as estimated by Sparrow
\cite{s82} in a different context. In Table~\ref{table1} there is only one
kneading pair $K=R^\infty$ and $H=L^\infty$ satisfying $H=L\overline{K}$. In
terms of the metric representations the condition (\ref{onep1}) defines a
straight line
(line {\bf a} in Fig.~\ref{hlzfig8})
\[ \alpha(H)=(3-\alpha(K))/4.\]
(We have used (\ref{alpha})). The point $r=28$ drops down to this line
almost vertically from the $r=36$ point. This is the region where ``fully
developed chaos'' has been observed in the Lorenz model and perhaps it
outlines the region where the geometric Lorenz model may apply.

The other limit (\ref{onep2}) happens at $r>197.6$. In Table~\ref{table1}
all~6 kneading pairs in this range satisfy $K=L\overline{H}$. They fall on
another straight line (line {\bf b} in Fig.~\ref{hlzfig8})
\[ \alpha(K)=(3-\alpha(H))/4,\]
but can hardly be resolved. The value $r=197.6$ manifests itself as the
point where the attractor no longer crosses the horizontal axis. In the
2D Poincar\'e map this is where the chaotic attractor stops to cross
the stable manifold ${\cal W}^s$ of the origin. The kneading pair at $r=203$
is very close to a limiting pair $K=LN(LR)^\infty$ with a precise value
$\alpha=21/40=0.525$ and $H=M(RL)^\infty$ with exact $\alpha=0.9$.

For any kneading pair in Table~\ref{table1} one can generate all admissible
periods up to length~6 inclusively. For example, at $r=125$ although the
swapped return map shown in Fig.~\ref{hlzfig5} exhibits some 2D feature as 
throwing a few points off the 1D attractor, the 1D Lorenz-Sparrow map still
works well. Besides the 17 orbits listed above for $r=118.15$, five new periods
appear: $LLN$, $LNLRMR$, $LNLLC$, and $RMRLN$. All these 22 unstable periodic
orbits have been located with high precision in the Lorenz equations. Moreover,
if we confine ourselves to short periods not exceeding period~6, then from
$r=28$ to 59.40 there are only symbolic sequences made of the two letters $R$
and $L$. In particular, From $r=28$ to 50.50 there exist the same 12 unstable
periods: $LR$, $RLR$, $RLLR$, $RRLR$, $RLRLR$, $RRLLR$, $RRRLR$, $RLRLLR$,
$RRLLLR$, $RRLRLR$, $RRRLLR$, and $RRRRLR$. This may partly explain the success
of the geometric Lorenz model leading to a symbolic dynamics on two letters. On
the other hand, when $r$ gets larger, e.g., $r=136.5$, many periodic
orbits ``admissible'' to the 1D Lorenz-Sparrow map, cannot be found in the
original Lorenz equations. This can only be analyzed by invoking 2D symbolic
dynamics of the Poincar\'e map.

\section*{V. SYMBOLIC DYNAMICS OF THE\protect\\
 2D POINCAR\'E MAPS}

\subsection*{Essentials of 2D symbolic dynamics}

The extension of symbolic dynamics from 1D to 2D maps is by no means trivial.
First of all, the 2D phase plane has to be partitioned in such a way as to
meet the requirements of a ``good'' partition that we put forward in Sec.~I.
Next, as 2D maps are in general invertible, a numerical orbit is encoded into
a {\em bi-infinite} symbolic sequence
\[ \cdots s_{\overline{m}} \cdots s_{\overline{2}} s_{\overline{1}} \bullet
 s_1 s_2 \cdots s_n \cdots, \]
where a heavy dot $\bullet$ denotes the ``present'' and one iteration
forward or backward corresponds to a left or right shift of the present dot.
The half-sequence $\bullet s_1s_2\cdots s_n \cdots$ is called a {\em forward}
symbolic sequence and $\cdots s_{\overline{m}}\cdots s_{\overline{2}}
 s_{\overline{1}}\bullet$ a {\em backward} symbolic sequence. One should
assign symbols to both forward and backward sequences in a consistent way by
partitioning the phase plane properly. In the context of the H\'enon map
Grassberger and Kantz \cite{gk85} proposed to draw the partition
line through tangent points between the stable and unstable manifolds of the
unstable fixed point in the attractor. Since pre-images and images of a tangent
point are also tangent points, it was suggested to take ``primary'' tangencies
where the sum of curvatures of the two manifolds is minimal \cite{gkm89}. 

A natural generalization of the Grassberger-Kantz idea is to use tangencies
between {\em forward contracting foliations} (FCFs) and {\em backward
contracting foliations} (BCFs) of the dynamics to determine the partition
line \cite{z9192}. Points on one and the same FCF approach each other with
the highest speed under forward iterations of the map. Therefore, one may
introduce an equivalence relation: points $p_1$ and $p_2$ belong to the same
FCF if they eventually approach the same destination under forward iterations
of the map:
\[ p_1 \sim p_2 \quad {\rm if} \quad
 \displaystyle\lim_{n\rightarrow\infty}|f^n(p_1)-f^n(p_2)|=0. \]
The collection of all FCFs forms a forward contracting manifold of the
dynamics. Points in one and the same FCF have the same future.

Likewise, points on one and the same BCF approach each other with the highest
speed under backward iterations of the map. One introduces an equivalence
relation: points $p_1$ and $p_2$ belong to the same BCF if they eventually
approach the same destination under backward iterations of the map:
\[ p_1 \sim p_2 \quad {\rm if} \quad
 \displaystyle\lim_{n\rightarrow\infty}|f^{-n}(p_1)-f^{-n}(p_2)|=0. \]
The collection of all BCFs forms the backward contracting manifold of the
dynamics. Points in one and the same BCF have the same history. When the phase
space is partitioned properly, points in a FCF acquire the same forward symbolic
sequence while points in a BCF acquire the same backward symbolic sequence.
this has been shown analytically for the Lozi map \cite{z9192} and T\'el map \cite{z92}. There has been good numerical evidence for the H\'enon map
\cite{cgp88,gkm89,zzg92,zz93}. We mention in passing that the forward
contracting and backward contracting manifolds contain the stable and unstable
manifolds of fixed and periodic points as invariant manifolds.

\begin{figure}
\centerline{\psfig{figure=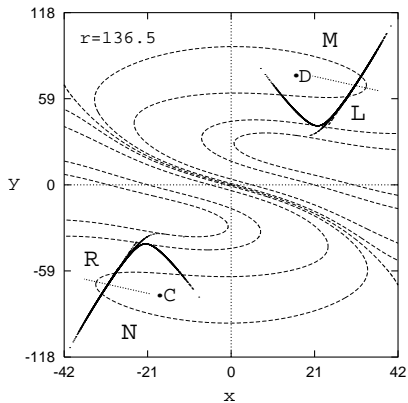,width=6cm,height=6cm}}
\caption{An upward Poincar\'e section at $r=136.5$ showing the chaotic
 attractor, a few FCFs (dashed lines), and segments of the partition lines for
 forward symbolic sequences (dotted lines).} 
\label{hlzfig9}
\end{figure}

The generalization to use FCFs and BCFs is necessary at least for the following
reasons. 1. It is not restricted to the attractor only. The attractor may
experience abrupt changes, but the FCF and BCF change smoothly with parameter.
This is a fact unproven but supported by much numerical evidence. 2. A good
symbolic dynamics assigns unique symbolic names to all unstable orbits, not
only those located in the attractor. One needs in partition lines outside the
attractor.  3. Transient processes also take place outside the attractor. They
are part of the dynamics and should be covered by the same symbolic dynamics. 

In practice, contours of BCFs and especially FCFs are not difficult to calculate
from the dynamics. This has been shown for BCFs by J. M. Greene \cite{g83} and
for FCFs by Gu \cite{gu87}. Once a mesh of BCFs and FCFs are drawn in the phase
plane, FCFs may be ordered along some BCF and {\it vice versa}. No ambiguity in
the ordering occurs as long as no tangency between the two foliations is
encountered. A tangency signals that one should change to a symbol of a
different parity after crossing a tangency. A tangency in a 2D map plays a role
similar to a kneading sequence in a 1D map in the sense that it prunes away some
inadmissible sequences. As there are infinitely many tangencies between the FCFs
and BCFs, one may say that there is an infinite number of kneading sequences in
a 2D map even at a fixed parameter set. However, as one deals with symbolic
sequences of finite length only a finite number of tangencies will matter.

\subsection*{Partitioning of the Poincar\'e section}

Lorenz equations at $r=136.5$ provide a typical situation where 2D symbolic
dynamics must be invoked. Figure~\ref{hlzfig9} shows an upward $z=r-1$
Poincar\'e section of the chaotic attractor. The dashed lines indicate the
contour of the FCFs. The two symmetrically located families of FCFs are
demarcated by the intersection of the stable manifold ${\cal W}^s$ of the
$(0, 0, 0)$ fixed point with the $z=r-1$ plane. The actual intersection
located between the dense dashed lines is not shown. The BCFs are not shown
either except the attractor itself which is a part of the BCFs. 

\begin{figure}
\centerline{\psfig{figure=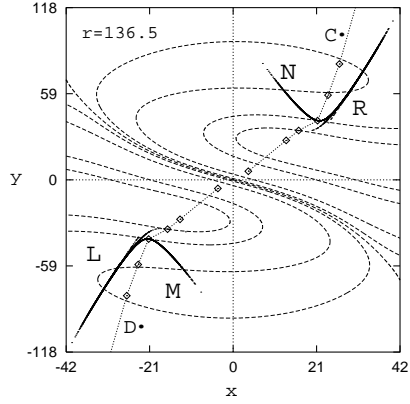,width=6cm,height=6cm}}
\caption{The same as Fig.~\ref{hlzfig9} showing 12 tangent points (diamonds)
 and the partition lines for backward symbolic sequences (dotted lines).}
\label{hlzfig10}
\end{figure}

The 1D symbolic dynamics analysis performed in Sec.~IV deals with forward
symbolic sequences only. However, the partition of the 1D interval shown e.g.,
in Fig.~\ref{hlzfig3}, may be traced back to the 2D Poincar\'e section to
indicate the partition for assigning symbols to the forward symbolic sequences.
Two segments of the partition lines are shown in Fig.~\ref{hlzfig9} as dotted
lines. The labels $\bullet C$ and $\bullet D$ correspond to $C$ and $D$ in
the Lorenz-Sparrow map, see Fig.~\ref{hlzfig3}. The ordering rule
(\ref{forder1}) should now be understood
as:
\begin{eqnarray*}
\bullet EN\cdots &<& \bullet EC\cdots < \bullet ER\cdots < \bullet
 EB\cdots\nonumber\\
 {} &<& \bullet EL\cdots < \bullet ED\cdots < \bullet EM\cdots,\nonumber\\
\bullet ON\cdots &>& \bullet OC\cdots > \bullet OR\cdots > \bullet
 OB\cdots\nonumber\\
 {} &>& \bullet OL\cdots > \bullet OD\cdots > \bullet OM\cdots,\nonumber\\
\end{eqnarray*}
with $E$ and $O$ being even and odd strings of $M$, $L$, $R$, and $N$. In fact,
from Fig.~\ref{hlzfig3} one could only determine the intersection point of the
partition line with the 1D-like attractor. To determine the partition line in
a larger region of the phase plane one has to locate more tangencies between
the FCFs and the BCFs. However, it is more convenient to use another set of
tangent points to determine the partition line for backward symbolic sequences.
To this end 6 tangent points and their mirror images are located and indicated
as diamonds in Fig.~\ref{hlzfig10}. The tangencies in the first quadrant are:
\begin{eqnarray*}
(3.833630661151,  5.915245399002),\\
(13.34721714210,  27.06932440906),\\
T_1: (16.50130604850,  33.81425621518),\\
T_4: (21.24012850767,  40.56850842796),\\
(23.86757424970,  58.00925911937),\\
(26.73829676387,  79.37583837912).\\
\end{eqnarray*}
(We have not labeled the tangent points off the attractor.)
The partition lines $C\bullet$ and $D\bullet$ are obtained by threading through
the diamonds. The two partition lines and the intersection with ${\cal W}^s$ of
the origin divide the phase plane into four regions, marked with the letters
$R$, $N$, $M$, and $L$. Among these 6 tangencies only $T_1$ and $T_4$ are
located on the attractor. Furthermore, they fall on two different sheets of the
attractor, making a 2D analysis necessary.

In order to decide admissibility of sufficiently long symbolic sequences
more tangencies on the attractor may be needed. These tangencies are taken
across the attractor. For example, on the partition line $C\bullet$ we have 
\begin{eqnarray}
\label{tangencies}
T_1:&  L^\infty RMC\bullet RRLRLLRLNRLLRLN \cdots\nonumber\\
{} & (16.501306048503,~ 33.814256215181) \nonumber\\
T_2:&  R^\infty RMC\bullet RRLLRRLLNRLRMLN \cdots\nonumber\\
{} & (16.567206430154,~ 34.823691929770)\nonumber\\
T_3:&  R^\infty RLC\bullet RLRMLRRLLNRLRML \cdots\nonumber\\
{} & (21.246853832518,~ 40.525036662442)\nonumber\\
T_4:&  L^\infty LLC\bullet RLRMLRRLRMRLLRL \cdots\nonumber\\
{} & (21.240128507672,~ 40.568508427961)\nonumber\\
\end{eqnarray}
Due to insufficient numerical resolution in Fig.~\ref{hlzfig10} the diamond on
the main sheet of the attractor represents $T_3$ and $T_4$, while the diamond
on the secondary sheet represents $T_1$ and $T_2$. The mirror images of these
tangencies are located on the $D\bullet$ partition line:
\begin{eqnarray*}
\bar T_1: &  R^\infty LND\bullet LLRLRRLRMLRRLRM \cdots \\
\bar T_2: &  L^\infty LND\bullet LLRRLLRRMLRLNRM \cdots\\
\bar T_3: &  L^\infty LRD\bullet LRLNRLLRRMLRLNR \cdots \\
\bar T_4: &  R^\infty RRD\bullet LRLNRLLRLNLRRLR \cdots \\
\end{eqnarray*}  

We denote the symbolic sequence of the tangency $T_i$ as $Q_iC\bullet K_i$,
keeping the same letter $K$ as the kneading sequence $K$ in the 1D
Lorenz-Sparrow map, because $K_i$ complies with the definition of a kneading
sequence as the next iterate of $C$. If one is interested in forward sequences
alone, only these $K_i$ will matter. Moreover, one may press together different
sheets seen in Fig.~\ref{hlzfig10} along the FCFs, as points on one and the
same FCF have the same forward symbolic sequence. Here lies a deep reason for
the success of 1D symbolic dynamics at least when only short periodic orbits
are concerned with. Therefore, before turning to the construction of 2D
symbolic dynamics let us first see what a 1D analysis would yield.

\subsection*{1D symbolic analysis at $r=136.5$}

\begin{figure}
\centerline{\psfig{figure=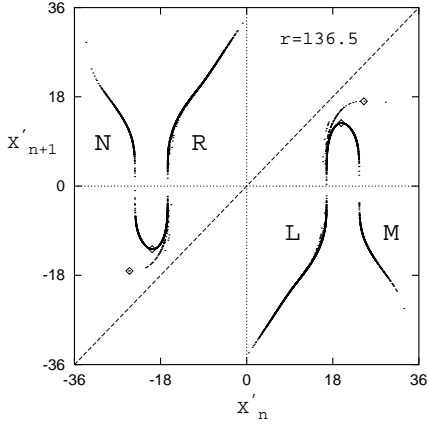,width=6cm,height=6cm}}
\caption{The swapped return map at $r=136.5$. }
\label{hlzfig11}
\end{figure}

Figure~\ref{hlzfig11} is a swapped return map obtained from the first return
map by letting the numerical constant be 41 in (\ref{swap}). The 2D feature
manifests itself as layers near $C$ and $D$. The four tangencies are plotted
as two diamonds in the figure, since $T_1$ is very close to $T_2$ and $T_3$
to $T_4$. As no layers can be seen away from the turning points one could only
get one $H$ from the set $\{|x_i^\prime|\}$. Now there are 4 kneading
sequences $K_i$, ordered as
\[ K_1 < K_2 < K_3 < K_4 \]
according to (\ref{forder1}). From the admissibility conditions (\ref{admis})
it follows that if two $K_- < K_+$ are both compatible with $H$, then any
symbolic sequence admissible under $(K_+, H)$ remains admissible under
$(K_-, H)$ but not the other way around. In our case, $(K_4, H)$ puts the
most severe restriction on admissibility while $(K_1, H)$ provides the weakest
condition. We start with the compatible kneading pair  
\begin{eqnarray}
\label{k1h}
K_1 & = & RRLRLLRLNRLLRLN\cdots,\nonumber\\
H & = & MNRRLLNLRRLRMLR\cdots.\nonumber\\
\end{eqnarray}
We produce all periodic symbolic sequences admissible under $(K_1, H)$ up to
period~6 using the procedure described in Appendix~B. The results are listed
in Table~\ref{table2}. Only shift-minimal sequences with respect to $N$ and $R$
are given. Their mirror images, i.e., shift-maximal sequences ending with $M$
or $L$, are also admissible.  There are in total 46 periods in
Table~\ref{table2}, where a $C$ stands for both $N$ and $R$. The kneading pair
$(K_2, H)$ forbids 2 from the 46 periods. The two pairs $(K_3, H)$ and
$(K_4, H)$ lying on the main sheet of the attractor have the same effect on
short periodic orbits. They reduce the allowed periods to 20, keeping those
from $LC$ to $RLRMLC$ in Table~\ref{table2}. The actual number of unstable
periodic orbits up to period~6 may be less than 46, but more than 20. A genuine
2D symbolic dynamics analysis is needed to clarify the situation.

\begin{table}
\caption{Admissible periodic sequences up to period~6 under the kneading pair
$(K_1, H)$ at $r=136.5$. Only the non-repeating shift-minimal strings with
respect to $N$ or $R$ are given. An asterisk marks those forbidden by 2D
tangencies, see text and Table~\ref{table3}.}
\label{table2}
\begin{center}
\begin{scriptsize}
\begin{tabular}{clcl}
\hline\hline
Period &  Sequence  &  Period & Sequence \\
\hline
2 & LC       &   6 & RLRRMR \\
4 & LNLR     &   3 & RLC$^*$ \\
6 & LNLRLC   &   6 & RLNRLR$^*$ \\
6 & LNLRMR   &   6 & RLNRMR \\
5 & LNLLC    &   5 & RLNLC$^*$ \\
3 & RMR      &   6 & RLNLLC$^*$ \\
5 & RMRLC    &   4 & RRMR \\
6 & RMLNLC   &   6 & RRMRLC$^*$ \\
4 & RMLN     &   6 & RRMRMR \\
4 & RLLC     &   5 & RRMLC$^*$ \\
6 & RLLNLC   &   6 & RRMLLN$^*$ \\
6 & RLRMLC   &   6 & RRLLLC$^*$ \\
6 & RLRLLC$^*$&  5 & RRLLC$^*$ \\
5 & RLRLC$^*$ &  6 & RRLRMC \\
\hline\hline
\end{tabular}
\end{scriptsize}
\end{center}
\end{table}

\subsection*{2D symbolic dynamics analysis at $r=136.5$}

In order to visualize the admissibility conditions imposed by a tangency
between FCF and BCF in the 2D phase plane we need metric representations
both for the forward and backward symbolic sequences. The metric representation
for the forward sequences remains the same as defined by (\ref{fmetric}).

The partition of phase plane shown in Fig.~\ref{hlzfig10} leads to a different
ordering rule for the backward symbolic sequences. Namely, we have
\[ L < D < M < B < N < C < R, \]
with the parity of symbols unchanged. The unchanged parity is related to the
positiveness of the Jacobian for the flow.) The ordering rule for backward
sequences may be written as:
\begin{eqnarray*}
\cdots LE\bullet < \cdots ME\bullet < \cdots NE\bullet < \cdots RE\bullet,\\
\cdots LO\bullet > \cdots MO\bullet > \cdots NO\bullet > \cdots RO\bullet,\\
\end{eqnarray*}
where $E$ ($O$) is a finite string containing an {\em even} ({\em odd}) number
of $M$ and $N$. From the ordering rule it follows that the maximal sequence
is $R^\infty\bullet$ and the minimal is $L^\infty\bullet$.
To introduce a metric representation for backward symbolic sequences, we
associate each backward sequence
$\cdots s_{\overline{m}} \cdots s_{\overline{2}}s_{\overline{1}}\bullet$
with a real number $\beta$:
\[ \beta= \sum_{i=1}^\infty \nu_{\overline i} 4^{-i}, \]
where
\[ \nu_{\overline i} = \cases{0\cr 1\cr 2\cr 3\cr} \quad {\rm for}
\quad s_i = \cases{L\cr M\cr N\cr R\cr}  \quad
 {\rm and }\quad \prod_{j=1}^{i-1}\epsilon_ {\overline j} = 1, \]
or
\[ \nu_{\overline i} = \cases{3\cr 2\cr 1\cr 0\cr} \quad {\rm for}
\quad s_i = \cases{L\cr M\cr N\cr R\cr}  \quad
 {\rm and }\quad \prod_{j=1}^{i-1}\epsilon_ {\overline j} = -1. \]
According to the definition we have
\begin{eqnarray*}
\beta(L^\infty \bullet) &=&0,\beta(D_\pm\bullet) =1/4,\beta(B_\pm\bullet)=1/2,\\
\beta (C_\pm \bullet) &=&3/4, \beta(R^\infty\bullet) =1,\\
\end{eqnarray*}

In terms of the two metric representations a bi-infinite symbolic sequence
with the present dot specified corresponds to a point in the unit square
spanned by $\alpha$ of the forward sequence and $\beta$ of the backward
sequence. This unit square is called the symbolic plane \cite{cgp88}.In the
symbolic plane forward and backward foliations become vertical and horizontal
lines, respectively. The symbolic plane is an image of the whole phase plane
under the given dynamics. Regions in the phase plane that have one and the
same forward or backward sequence map into a vertical or horizontal line in
the symbolic plane. The symbolic plane should not be confused with the
$\alpha(H) \sim \alpha(K)$ plane (Fig.~\ref{hlzfig8}) which is the metric
representation of the kneading plane, i.e., the parameter plane of a 1D map.

\begin{figure}
\centerline{\psfig{figure=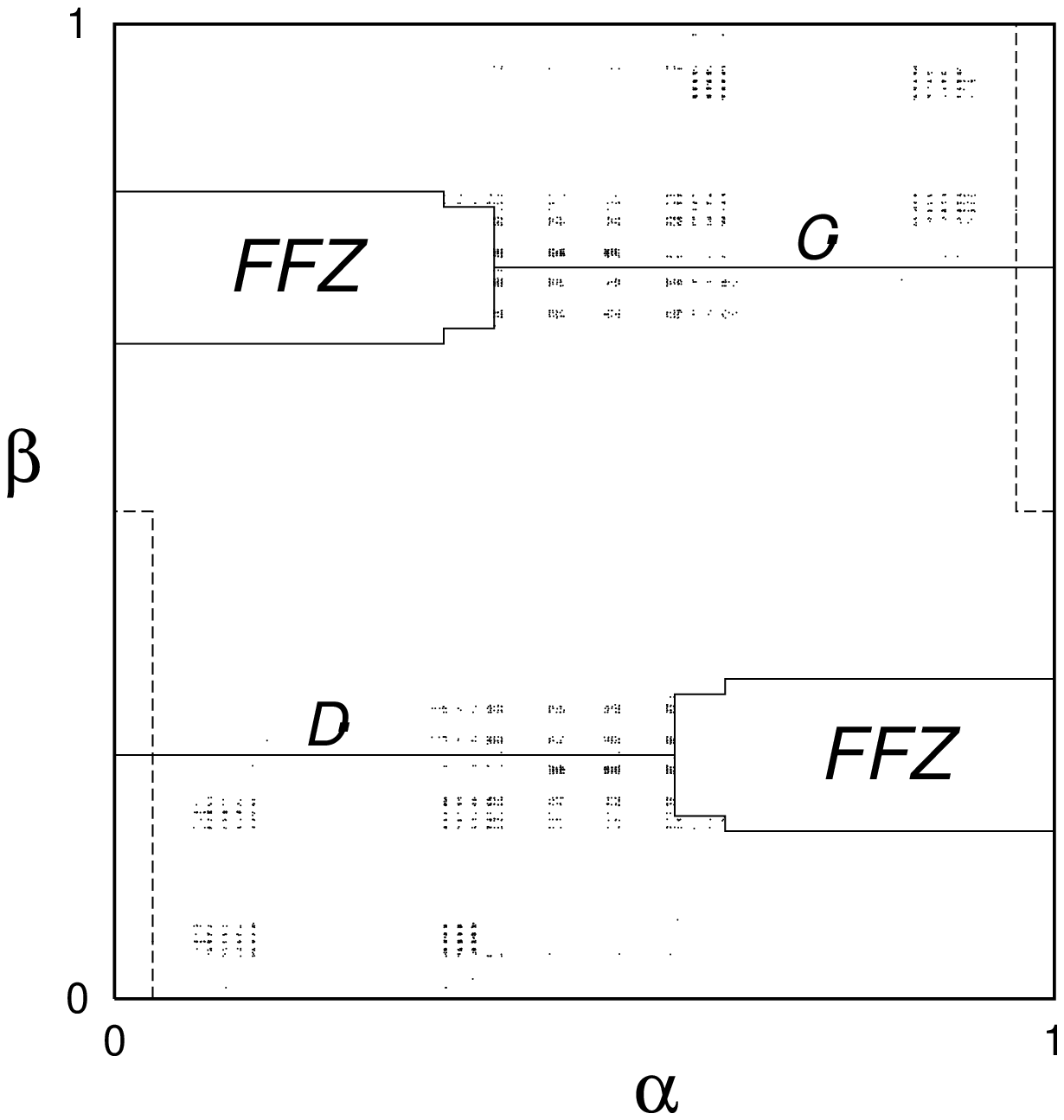,width=6cm,height=6cm}}
\caption{The symbolic plane at $r=136.5$. A total of 60000 points representing
real orbits are drawn together with the FFZ outlined by the 4 tangencies and
the forbidden zone caused by $H$.}
\label{hlzfig12}
\end{figure}

As long as foliations, i.e., symbolic sequences, are well ordered, a tangency
on a partition line puts a restriction on allowed symbolic sequences.
Suppose that there is a tangency $QC\bullet K$ on the partition line $C\bullet$.
The rectangle enclosed by the lines $\beta(QR\bullet)$, $\beta(QN\bullet)$,
$\alpha(\bullet K)$, and $\alpha((MN)^\infty)=0$ forms a forbidden zone (FZ) in
the symbolic plane. In the symbolic plane a forbidden sequence corresponds to a
point inside the FZ of $QC\bullet K$. A tangency may define some allowed
zones as well. However, in order to confirm the admissibility of a sequence
all of its shifts must locate in the allowed zones, while one point in the FZ
is enough to exclude a sequence. This ``all or none'' alternative tells us
that it is easier to exclude than to confirm a sequence by a single tangency.
Similarly, a tangency $\overline{Q}D\bullet \overline{K}$ on the partition line 
$D\bullet$ determines another FZ, symmetrically located to the FZ mentioned
above. Due to the anti-symmetry of the map one may confine oneself to the
first FZ and to shift-maximal sequences ending with $N$ and $R$ only when
dealing with finite periodic sequences. The union of FZs from all possible
tangencies forms a fundamental forbidden zone (FFZ) in the $\alpha - \beta$
symbolic plane. A necessary and sufficient condition for a sequence to be
allowed consists in that all of its shifts do not fall in the FFZ. Usually,
a finite number of tangencies may produce a fairly good contour of the
FFZ for checking the admissibility of finite sequences. In Fig.~\ref{hlzfig12}
we have drawn a symbolic plane with $60000$ points representing real orbits
generated from the Poincar\'e map at $r=136.5$ together with a FFZ, outlined
by the four tangencies (\ref{tangencies}). The other kneading sequence $H$ in
the 1D Lorenz-Sparrow map bounds the range of the 1D attractor. In the 2D
Poincar\'e map the sequence $H$ corresponds to the stable manifold of the
origin which intersects with the attractor and bounds the subsequences
following an $R$. In the symbolic plane the rectangle formed by
$\alpha=\alpha(H)$, $\alpha=1$, $\beta=0.5$, and $\beta=1$ determines the
forbidden zone caused by $H$. It is shown in Fig.~\ref{hlzfig12} by dashed
lines. Indeed, the FFZ contains no point of allowed sequences.

\begin{table}
\caption{Location of admissible periodic orbits left from Table~\ref{table2}
by 2D analysis. The coordinates $(x, y)$ are that of the first symbol in a
sequence.}
\label{table3}
\begin{center}
\begin{scriptsize}
\begin{tabular}{clcc}
\hline\hline
Period & \multicolumn{1}{c}{Sequence} &\multicolumn{1}{c}{$x$} &
 \multicolumn{1}{c}{$y$} \\
\hline
   2 &       LR      & -26.789945953 & -51.732394996\\
   2 &       LN      & -33.741639204 & -79.398248620 \\
   4 &    LNLR       & -34.969308137 & -84.807257714\\
   6 &  LNLRLN       & -34.995509382 & -84.923968314\\
   6 &  LNLRLR       & -35.378366481 & -86.639695512\\
   6 & LNLRMR        & -36.614469777 & -92.269654794\\
   5 &  LNLLN        & -36.694480374 & -92.638862207\\
   5 &  LNLLR        & -37.362562975 & -95.744924312\\
   3 &  RMR          &  36.628892834 & 92.335783415\\
   5 &  RMRLN        &  36.548092868 & 91.963380870\\
   5 &  RMRLR        &  35.927763416 & 89.123769188\\
   6 & RMLNLR        &  33.541019900 & 78.514904719\\
   6 & RMLNLN        &  33.465475168 & 78.187866239\\
   4 & RMLN          &  33.432729468 & 78.045902429\\
   4 & RLLR          &  29.500017415 & 61.545331390\\
   4 & RLLN          &  28.800493901 & 58.709002527\\
 6  & RLLNLN         &  28.566126025 & 57.759480656 \\
 6  & RLLNLR         &  28.548686604 & 57.682625650 \\
  6 & RLRMLN         &  28.310187611 & 56.723089485\\
  6 & RLRMLR         &  28.299181162 & 56.676646246\\
  6 & RLRRMR         &  26.376239173 & 48.942140325\\
  6 & RLNRMR         &  25.282520197 & 44.566367330\\
  4 & RRMR           &  25.163031306 & 44.088645613\\
  6 & RRMRMR         &  25.047268287 & 43.625877472\\
  6 & RRLRMN         &  24.055406683 & 39.708285078\\
  6 & RRLRMR         &  24.064390385 & 39.704320864\\
\hline\hline
\end{tabular}
\end{scriptsize}
\end{center}
\end{table}

In order to check the admissibility of a period~$n$ sequence one calculates
$n$ points in the symbolic plane by taking the cyclic shifts of the
non-repeating string. All symbolic sequences listed in Table~\ref{table2}
have been checked in this way and 20 out of 46 words are forbidden, in fact,
by $T_3$. This means among the 26 sequences forbidden by $K_4$ in a 1D analysis
actually 6 are allowed in 2D. We list all admissible periodic sequences
of length~6 and less in Table~\ref{table3}. The 6 words at the bottom of the
table are those forbidden by 1D but allowed in 2D. All the unstable periodic
orbits listed in Table~\ref{table3} have been located with high precision in
the Lorenz equations. The knowledge of symbolic names and the ordering rule
significantly facilitates the numerical work. The coordinates $(x, y)$ of the
first symbol of each sequence are also given in Table~\ref{table3}.

\subsection*{Chaotic orbits}

Symbolic sequences that correspond to chaotic orbits also obey the ordering
rule and admissibility conditions. However, by the very definition these
sequences cannot be exhaustively enumerated. Nevertheless, it is possible to
show the existence of some chaotic symbolic sequences in a constructive way.

We first state a proposition similar to the one mentioned in the paragraph
before Eq.~(\ref{k1h}). If $(K, H_-)$ and $(K, H_+)$ are two compatible
kneading pairs with $H_- < H_+$, then all admissible sequences under
$(K, H_-)$ remain so under $(K, H_+)$, but not the other way around. It is
seen from Table~\ref{table1} that from $r=120.0$ to 191.0 all $K$
starts with $R$ while the minimal $H$ starts with $MRR$. Let $K=R\cdots$
and $H=MRR\cdots$. It is easy to check that any sequence made of the two
segments $LR$ and $LNLR$ satisfies the admissibility conditions(\ref{admis}).
Therefore, a random combination of these segments is an admissible sequence
in the 1D Lorenz-Sparrow map. A similar analysis can be carried out in 2D
using the above tangencies. Any combination of the two segments remains an
admissible sequence in 2D. Therefore, we have indicated the structure of ai
class of chaotic orbits in th eparameter range.

\section*{VI. STABLE PERIODIC ORBITS IN\protect\\
 LORENZ EQUATIONS} 

So far we have only considered unstable periodic orbits at fixed $r$. A good
symbolic dynamics should be capable to deal with stable orbits as well.
One can generate all compatible kneading pairs of the Lorenz-Sparrow map
up to a certain length by using the method described in Appendix~A.
Although there is no way to tell the precise parameter where a given periodic
orbit will become stable, the symbolic sequence does obey the ordering rule
and may be located on the $r$-axis  by using a bisection method. Another way
of finding a stable period is to follow the unstable orbit of the same name
at varying parameters by using a periodic orbit tracking program. Anyway, many
periodic windows have been known before or encountered during the
present study. We collect them in Tables~\ref{table4} and~\ref{table5}.
Before making further remarks on these tables, we indicate how to find
symbolic sequences for stable periods.

When there exists a periodic window in some parameter range, one
cannot extract a return map of the interval from a small number of orbital
points so there may be ambiguity in assigning symbols to numerically
determined orbital points. Nonetheless, there are at least two ways to
circumvent the difficulty. First, one can take a nearby parameter where the
system exhibits chaotic behavior and superimpose the periodic points on the
chaotic attractor. In most cases the $(K, H)$ pair calculated from the chaotic
attractor may be used to generate unstable periods coexisting with the
stable period. Second, one can start with a set of initial points and keep as
many as possible transient points before the motion settles down to the final
stable periodic regime (a few points near the randomly chosen initial points
have to be dropped anyway). From the set of transient points one can construct
return maps as before. Both methods work well for short enough periods,
especially in narrow windows.

\begin{figure}
\centerline{\psfig{figure=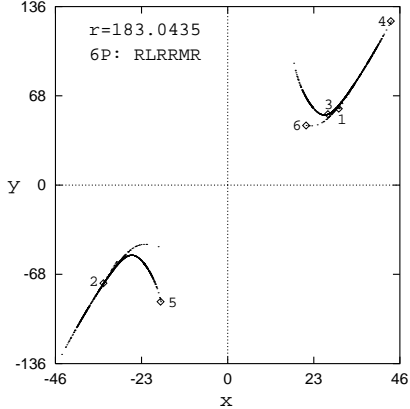,width=6cm,height=6cm}}
\caption{A stable period~6 orbit at $r=183.0435$ on the background of its own
transient points.}
\label{hlzfig13}
\end{figure}

Fig.~\ref{hlzfig13} shows a stable period~6 orbit $RLRRMR$ at $r=183.0435$
as diamonds. The background figure looks much like a chaotic attractor, but
it is actually a collection of its own transient points. The last symbol $X$
in $RLRRMX$ corresponds to a point $(20.945669, 45.391029)$ lying to the right
of a tangency at $(20.935971, 45.393162)$. Therefore, it acquires the symbol
$R$, not $N$. This example shows once more how the $x$-parameterization helps
in accurate assignment of symbols.

\subsection*{Absolute nomenclature of periodic orbits}
 
In a periodically driven system the period of the external force serves as a
unit to measure other periods in the system. This is not the case in autonomous
systems like the Lorenz equations, since the fundamental frequency drifts with
the varying parameter. No wonder in several hundred papers on the Lorenz model
no authors had ever described a period as, say, period~5 until a calibration
curve of the fundamental frequency was obtained by extensive Fourier analysis
in \cite{dhh85}. It is remarkable that the absolute periods thus obtained
coincide with that determined later from symbolic dynamics (\cite{dh88,fh96}
and the present paper). As a consequence, we know now that the window first
studied by Manneville and Pomeau \cite{mp79} starts with period~4, the
period-doubling cascade first discovered by Franceschini \cite{f80} lives in
a period~3 window, etc. Moreover, we know their symbolic names and their
location in the overall systematics of all stable periods. In
Tables~\ref{table4} and~\ref{table5} there are many period-doubled
sequences, whose numerically determined symbolic names all comply with the
rules of symbolic dynamics. 

\subsection*{Symmetry breakings and restorations}

In a dynamical system with discrete symmetry the phenomenon of symmetry
breaking and symmetry restoration comes into play. In the Lorenz equations
periodic orbits are either symmetric or asymmetric with respect to the
transformation (\ref{invers}); asymmetric orbits appear in symmetrically located
pairs. Some essential features of symmetry breaking and restoration have been
known. For example, symmetry breaking must precede period-doubling --- no
symmetric orbits can experience period-doubling directly without the symmetry
being broken first; symmetry breakings take place in periodic regime, and
symmetry restorations occur in chaotic regime etc. All these features may be
explained by using symbolic dynamics \cite{zh89}. Although the analysis
performed in \cite{zh89} was based on the anti-symmetric cubic map, it is
applicable to the Lorenz equations via the Lorenz-Sparrow map.

A doubly ``superstable'' symmetric orbit must be of the form $\Sigma D
\overline{\Sigma} C$, therefore its period is even and only even periods of
this special form may undergo symmetry breaking. The shortest such orbit is
$DC$.  To keep the symmetry when extending this superstable period into a
window, one must change $D$ and $C$ in a symmetric fashion, i.e., either
replacing $D$ by $M$ and $C$ by $N$ or replacing $D$ by $L$ and $C$ by $R$ at
the same time, see (\ref{order1}) and (\ref{trans}). Thus we get a window
$(MN, DC, LR)$ ($MN$ does not appear in the Lorenz equations while $LR$
persists to very large $r$). This is indeed a symmetric window, as the
transformation (\ref{trans}) brings it back after cyclic permutations.
Moreover, this window has a signature $(+, 0, +)$ according to the parity of
the symbols (we assign a null parity to $C$ and $D$). It cannot undergo
period-doubling as the latter requires a $(+, 0, -)$ signature. By continuity
$LR$ extends to an asymmetric window $(LR, LC, LN)$ with signature $(+, 0, -)$
allowing for period-doubling. It is an asymmetric window as its mirror image
$(RL, RD, RM)$ is different. They represent the two symmetrically located
asymmetric period~2 orbits. The word $(LR)^\infty$ describes both the second
half of the symmetric window and the first half of the asymmetric window. The
precise symmetry breaking point, however, depends on the mapping function and
cannot be told by symbolic dynamics.

In general, a word $\lambda^\infty$ representing the second half of the
symmetric window continues to become the first half of the asymmetric window
$(\lambda, \tau C, \rho)$. The latter develops into a period-doubling cascade
described by the general rule of symbolic dynamics. The cascade accumulates and
turns into a period-halving cascade of chaotic bands. The whole structure is
asymmetric. Finally, the chaotic attractor collides with the symmetric unstable
periodic orbit $\lambda^\infty$ and takes back the symmetry to become a
symmetric chaotic attractor. This is a symmetry restoration crisis, taking
place at the limit described by the eventually periodic kneading sequence
$\rho\lambda^\infty$. In our period~2 example this happens at $LN(LR)^\infty$.
In Table~\ref{table4} this limit has been traced by $LN(LR)^{n-2}LC$ up to
$n=15$. The only period~{30} sequence in Table~\ref{table4} indicates closely
the location of the symmetry restoration point corresponding to the asymmetric
period~$2^n$ cascade. All other symmetric orbits in Table~\ref{table4} are put
conditionally in the form $\Sigma D\overline{\Sigma}C$ as the parameters given
can hardly match a doubly superstable orbit. For example, the three consecutive
period~4 from $r=148.2$ to 166.07 in Table~\ref{table4} actually mean:
\begin{eqnarray*}
(RMLN, RDLC, RLLR) &\rightarrow& (RLLR, RLLC, RLLN),\\
(+, 0, +) &\rightarrow& (+, 0, -),\\
\end{eqnarray*}
followed by an asymmetric period-doubling cascade. The symmetry restores at
$RLLN(RLLR)^\infty$ whose parameter may easily be estimated.

Symbolic dynamics also yields the number of periodic orbits that are capable
to undergo symmetry breaking. In the parameter range of the Lorenz equations
there are one period~2, one period~4, and two period~6 such orbits, all listed
in Table~\ref{table4}.

\subsection*{``2D'' orbits and co-existing attractors}

Now we return to Tables~\ref{table4} and~\ref{table5}. Table~\ref{table4}
is a list of stable periods associated with the main sheet of the dynamical
foliations. When there is an attracting stable period these sheets are not
readily seen, but they resemble the main sheets seen in Fig.~\ref{hlzfig5} or
Fig.~\ref{hlzfig11}. In fact, one may insert all the kneading sequences $K_i$
listed in Table~\ref{table1} into Table~\ref{table4} according to their
$r$-values. They all fit well into the overall ordering. The ordered list of
stable periods plus that of kneading sequence $K_i$ determined from the main
sheets of the chaotic attractor makes an analogue of the MSS-sequence
\cite{mss} in the symbolic dynamics of unimodal maps. It is a surprising fact
the 1D Lorenz-Sparrow map captures so much of the real Lorenz equations.
Then where are the manifestly 2D features? As long as stable periods are
concerned, some orbits showing 2D features are collected in Table~\ref{table5}.
As a rule, these are very narrow windows living on some secondary sheets of the
dynamical foliations. It is remarkable that they may be named according to the
same rule of the Lorenz-Sparrow map; they form a different ordered list as
compared with Table~\ref{table4}. Among them there are a few orbits co-existing
with a periodic orbit from the main sheet especially when the latter forms a
wide window. For instance, $RMN$ and $LR$ coexist in the vicinity of
$r=328.0838$. This period~3 orbit develops a period-doubling cascade, traced to
period~{24} in Table~\ref{table5}. The period~2 orbit $(LR)^\infty$ may even be
seen co-existing with a tiny chaotic attractor from the same $3^n$ cascade at
$r=327.16755$. Other cases, given in Table~\ref{table5}, include $RRMR$ and
$RRMRMC$ as well as their period-doubled regimes, both coexisting with the
symmetric period~4 orbit $RLLR$ below $r=162.1381$ and 157.671066,
respectively. In addition, there are orbits involving both sheets. We attribute
all these orbits to the manifestation of 2D features.

\section*{VII. CONCLUDING REMARK}

Fairly detailed global knowledge of the Lorenz equations in the phase space as
well as in the parameter space has been obtained by numerical work under the
guidance of symbolic dynamics. Two-dimensional symbolic dynamics of the
Poincar\'e map may provide, in principle, a complete list  of stable and
unstable periodic orbits up to a given length and a partial description of
some chaotic orbits. However, 1D symbolic dynamics extracted from the
2D Poincar\'e map is simpler and instructive. The 2D features seen in the 
Poincar\'e and first return maps may safely be circumvented by shrinking
along the FCFs in a 1D study which deals with forward symbolic sequences
only. Whether 1D or 2D symbolic dynamics is needed and how many tangencies
to take in a 2D study is a matter of precision. Even in a seemingly ``pure''
one-dimensional situation 2D features may need to be taken into account when it
comes to cope with very long symbolic sequences. This has to be decided in
practice. Therefore, what has been described in this paper remains a
physicist's approach for the time being. However, it may provide food for
thought to mathematicians. We mention in passing that there are some
technical subtleties in carrying out the program that we could not touch upon
due to limited space, but there is also a good hope to automate the process
and to apply it to more systems of physical importance.

\section*{ACKNOWLEDGMENTS}

This work was partly supported by the Chinese NSF and Nonlinear Science
Project. Discussions with Drs. H.-P. Fang, Z.-B. Wu, and F.-G. Xie are
gratefully acknowledged. BLH thanks Prof. H. C. Lee and the National Central
University in Chung-li, Taiwan, and JXL thanks Prof. M. Robnik and the Center
for Applied Mathematics and Theoretical Physics at University of Maribor,
Slovenia, for the hospitality while the final version of this paper was written.

\section*{Appendix A. Generation of compatible kneading \protect\\
 pairs for the Lorenz-Sparrow map}

We first make a proviso relevant to both Appendices. In a 2D setting it is
difficult to say that a symbolic sequence is included ``between'' some other
sequences without further specifying how the order is defined. In a phase space
a number of FCFs may be ordered along a BCF that intersects with some FCFs
transversely. In the parameter space of the Lorenz equations we are working
along the $r$-axis and there is a 1D ordering of all symbolic sequences
according to the Lorenz-Sparrow map. We hope this is clear from the context in
what follows.

Two kneading sequences $K$ and $H$ must satisfy the admissibility conditions
(\ref{admis}) in order to become a compatible kneading pair $(K, H)$. This
means, in particular, two sequences with $H<K$ cannot make a compatible
pair. Moreover, from the admissibility conditions one can deduce that
the minimal $H$ that is compatible with a given $K$ is given by
\[ K\leq H_{\rm min}={\rm max}\{ {\cal R}(K),{\cal R}(\overline{K}),
 {\cal N}(K),{\cal N}(\overline{K})\},\]
where ${\cal R}(K)$ is the $R$-shift set of $K$, etc.

According to the ordering rule (\ref{forder1}) the greatest sequence is
$(MN)^\infty$, and the smallest $(NM)^\infty$. A sequence $H=(MN)^\infty$ will
be compatible with any $K$. The admissibility conditions also require
that $K$ must be shift-minimal with respect to $R$ and $N$. Both $(MN)^\infty$
and $(NM)^\infty$ meet this requirement. Taking the extreme sequences
$K_1=(MN)^\infty$, $K_2=(NM)^\infty$ and $H=(MN)^\infty$, one can
generate all compatible kneading pairs up to a certain length by making use
of the following propositions (in what follows $\Sigma=s_0s_1\cdots s_n$
denotes a finite string of $M$, $L$, $R$, and $N$, and
$\mu, \nu\in\{M, L, R, N\}$, $\mu\neq \nu$):\\
\indent 1. If $K_1=\Sigma\mu\cdots$ and $K_2=\Sigma\nu\cdots$ are both
compatible with a given $H$, then $K=\Sigma\tau$ is also compatible with
$H$, where $\tau\in\{C, B, D\}$ is included between $\mu$ and $\nu$,
i.e., either $\nu<\tau<\mu$ or $\nu>\tau>\mu$ holds.

2. For $\tau=C$, under the conditions of 1, $K=(\Sigma t)^\infty$ is
also compatible with $H$, where $t$ stands for either $R$ or $N$.

3. For $\tau=D$, under the conditions of 1, $K=(\Sigma t\overline{\Sigma}
\overline{t})^\infty$ is compatible with $H$, where $t$ stands for either
$M$ or $L$.

4. For $\tau=B$, under the conditions of 1, $K_1=\Sigma RH$ and
$K_2=\Sigma L\overline{H}$ are both compatible with $H$.

Without going into the proofs we continue with the construction. By means of the
above propositions we have the median words $K= D, B, C$ between $K_1$ and
$K_2$. At this step we have the following words, listed in ascending order:
\[(NM)^\infty\quad N^\infty,C,R^\infty\quad R(MN)^\infty,B,L(NM)^\infty\quad (LR)^\infty,DC,(MN)^\infty\quad (MN)^\infty. \]
Inside any group centered at $C$, $B$, or $D$ there exists no median sequence.
Furthermore, no median sequence exists between the group $D$ and $(MN)^\infty$.
Taking any two nearby different sequences between the groups, the procedure may
be continued. For example, between $R^\infty$ and $R(MN)^\infty$ we get
\[ R^\infty\quad RR(MN)^\infty,RB,RL(NM)^\infty\quad (RLLR)^\infty,RDLC,(RMLN)^\infty\quad R(MN)^\infty.\]
This process is repeated to produce all possible $K$ up to a certain length.
For each $K$ one determines a $H_{\rm min}$. In this way we construct the
entire kneading plane for the Lorenz-Sparrow map. Fig.~\ref{hlzfig8} shows
that only a small part of this plane is related to the Lorenz equations.
This is caused mainly by the set of $H$ that may occur in the system. 

The above method may be applied to the Lorenz equations to generate and locate
median words included between two known stable orbits.
For example, between $RRRLC$ at $r^\prime_1=59.247$ and $RRRRLLC$ at
$r^\prime_2=55.787$ two period~9 words $RRRRLLLLC$ and $RRRRLLRLC$ can be
produced as follows. At first, take $r_1=59.40$ and $r_2=55.90$ near the two
$r^\prime$-values and determine the corresponding maximal sequences $H$ from
the chaotic attractors as we did above at $r=181.15$ or 136.5. They turn out
to be $H_1=MLLLRLLRL\cdots$ for the former and $H_2=MLLLLRRLR\cdots $ for the
latter. Then take their common string to be a new $H=MLLL$. Finally we can
use $K_1=RRRLC$, $K_2=RRRRLLC$, and $H=MLLL$ to form compatible kneading pairs 
and to generate $RRRRLLLLC$ and $RRRRLLRLC$ which are included in between
$RRRLC$ and $RRRRLLC$. In order to have this procedure working well, the
difference $|r_1 - r_2|$ should be small to guarantee that $H$ is long enough
to be usable. In addition, $r_1$ and $r_2$ should be chosen close enough to
$r^\prime_1$ and $r^\prime_2$ so that $K_{r_1} = RRRLRRRRLRRRLLL\cdots$ at
$r_1$ and $K_{r_2} = RRRRLLRRRRLRLLR\cdots$ at $r_2$ are very close to $RRRLC$
and $RRRRLLC$, respectively.

\section*{Appendix B. Generation of admissible sequences\protect\\
 for a given kneading pair}

Given a compatible kneading pair $(K, H)$, one can generate all admissible
symbolic sequences up to a given length, e.g.,~6. Usually, we are interested
in having a list of symbolic names of all short unstable periodic orbits.
This can be done by brute force, i.e., first generate all $6^4$ possible
symbolic sequences then filter them against the admissibility conditions
(\ref{admis}). In so doing one should avoid repeated counting of words.
Therefore, we always write the basic string of a periodic sequence in the
shift-minimal form with respect to $N$ or $R$. The shift-maximal sequences
with respect to $M$ or $L$ may be obtained by applying the symmetry
transformation ${\cal T}$.

However, one can formulate a few rules to generate only the admissible
sequences. These rules are based on continuity in the phase plane. To
simplify the writing we introduce some notation. Let $\Sigma_n=s_1 s_2
\cdots s_n$ be a finite string of $n$ symbols; let symbols $\mu$, $\nu$,
and $s_i,\;\; i=1, 2, \cdots, n$ be all taken from the set $\{M, L, R, N\}$;
and let the symbol $\tau$ denote one of $\{C, B, D\}$. Recollect, moreover,
that at any step of applying the rules a $C$ at the end of a string is to be
continued as $CK$, a $D$ as $D\overline{K}$, and a $B$ as $RH$ or
$L\overline{H}$, see (\ref{knead}).

We have the following propositions:\\
\indent 1. If both $\Sigma_n\mu\cdots$ and $\Sigma_n\nu\cdots$ are admissible,
 then $\Sigma_n\tau\cdots$ is admissible provided $\tau$ is include
 between $\mu$ and $\nu$, i.e., either $\nu < \tau < \mu$ or $\nu > \tau >
 \mu$ takes place.

2. If $\Sigma_n B$ and $\Sigma_n C$ are admissible then so does
 $(\Sigma_n R)^\infty$.
  
3. If $\Sigma_n C$ and $\Sigma_n\mu\cdots$ are admissible and, in addition,
 $\Sigma_n t K< (\Sigma_n t)^\infty < \Sigma_n\mu\cdots$, where $t\in
 \{R, N\}$, then $(\Sigma_n t)^\infty$ is admissible.

4. If $\Sigma_n\mu\cdots$, $\Sigma_n D$ and $\Sigma_n\nu\cdots$ are admissible,
 $\Sigma_n t$ and $\Sigma_n w$ are respectively the greater and the smaller of
 $\Sigma_n L$ and $\Sigma_n M$, then
 $\Sigma_n t\overline{K}< (\Sigma_n t\overline{\Sigma_n}w)^\infty <
 \Sigma_n\mu$ implies the admissibility of
 $(\Sigma_n t\overline{\Sigma_n}w)^\infty$ or
 $\Sigma_n\nu\cdots < (\Sigma_n w\overline{\Sigma_n}t)^\infty <
 \Sigma_n w\overline{K}$ implies the admissibility of
 $(\Sigma_n w\overline{\Sigma_n}t)^\infty$.

5. If $I_1=u_1u_2\cdots u_nB\equiv UB$ and $I_2=U R\cdots$ are admissible
 and the leading string of $I_2$ turns out to be $u_1u_2\cdots u_k
 \tau$ with $k<n$ and $\tau\in\{C, B, D\}$ , then $(UR)^\infty$ is admissible
 if it is included between $I_1$ and $I_2$, otherwise it is inadmissible.
 Similarly, If $I_1=UB$ and $I_2=UL\cdots$ are admissible, then
 $(UL\overline{U}R)^\infty$ is admissible if it is included between
 $I_1$ and $I_2$, otherwise it is inadmissible.

We omit the proofs \cite{z97} of these propositions which are based on
continuity in the phase plane and on explicit checking of the admissibility
conditions (\ref{admis}).

\begin{table}
\caption{Some stable periodic orbits in the Lorenz equations associated with
the main sheet of the dynamical foliations.}
\label{table4}
\begin{center}
\begin{scriptsize}
\begin{tabular}{llc}
\hline\hline
\multicolumn{1}{c}{Period} & \multicolumn{1}{c}{Sequence} & $r$ \\
\hline
   2          & DC                          & 315-10000 \\
   2          & LN                          & 229.42-314  \\
  $\;\;\;$4   & LNLC                        & 218.3-229.42 \\
  $\;\;\;$8   & LNLRLNLC                    & 216.0-218.3  \\
$\;\;$16      & LNLRLNLNLNLRLNLC            & 215.5-216.0 \\
 24           & LNLRLNLNLNLRLNLRLNLRLNLC    & 215.07-215.08 \\
 12           & LNLRLNLNLNLC                & 213.99-214.06  \\
  6           & LNLRLC                      & 209.06-209.45 \\
$\;\;$12      & LNLRLRLNLRLC                & 208.98  \\
 10           & LNLRLRLNLC                  & 207.106-207.12 \\
  8           & LNLRLRLC                    & 205.486-206.528  \\
 10           & LNLRLRLRLC                  & 204.116-204.123 \\
 12           & LNLRLRLRLRLC                & 203.537  \\
 14           & LNLRLRLRLRLRLC              & 203.2735  \\
 16           & LNLRLRLRLRLRLRLC            & 203.1511  \\
 18           & LNLRLRLRLRLRLRLRLC          & 203.093332 \\
 30           & LNLRLRLRLRLRLRLRLRLRLRLRLRLRLC & 203.04120367965  \\
 14           & LNLRLRDRMRLRLC              & 200.638-200.665\\
 10           & LNLRDRMRLC                  & 198.97-198.99  \\
  5           & LNLLC                       & 195.576  \\
$\;\;$10      & LNLLRLNLLC                  & 195.564  \\
$\;\;$20      & LNLLRLNLLNLNLLRLNLLC        & 195.561  \\
  5           & RMRLC                       & 190.80-190.81 \\
 $\;\;$10     & RMRLRRMRLC                  & 190.79  \\
$\;\;$20      & RMRLRRMRLNRMRLRRMRLC        & 190.785  \\
  7           & RMRLRLC                     & 189.559-189.561 \\
  9           & RMRLRLRLC                   & 188.863-188.865 \\
 16           & RMRLRLRDLNLRLRLC            & 187.248-187.25 \\
 12           & RMRLRDLNLRLC                & 185.74-185.80 \\
  8           & RMRDLNLC                    & 181.12-181.65 \\
 10           & RMRMRMLNLC                  & 178.0745 \\
 12           & RMRMRDLNLNLC                & 177.78-177.81 \\
 6            & RMLNLC                      & 172.758-172.797 \\
$\;\;$12      & RMLNLNRMLNLC                & 172.74 \\
 16           & RMLNRMRDLNRMLNLC            & 169.902 \\
 10           & RMLNRMLNLC                  & 168.58 \\
  4           & RDLC                        & 162.1-166.07 \\
  4           & RLLC                        & 154.4-162.0 \\
  4           & RLLN                        & 148.2-154.4 \\
 $\;\;\;$8    & RLLNRLLC                    & 147.4-147.8  \\
$\;\;$16      & RLLNRLLRRLLNRLLC            & 147 \\
 12           & RLLNRLLRRLLC                & 145.94-146  \\
 20           & RLLNRLLRRDLRRMLRRLLC        & 144.35-144.38 \\
 12           & RLLNRDLRRMLC                & 143.322-143.442\\
 6            & RLLNLC                      & 141.247-141.249 \\
$\;\;$12      & RLLNLRRLLNLC                & 141.23  \\
  6           & RLRMLC                      & 136.79-136.819 \\
 $\;\;\;$12   & RLRMLRRLRMLC                & 136.795  \\
 10           & RLRMLRRLLC                  & 136.210-136.2112 \\
 16           & RLRMLRRDLRLNRLLC            & 135.465-135.485\\
  8           & RLRDLRLC                    & 132.06-133.2 \\
 16           & RLRLLRRDLRLRRLLC            & 129.127-129.148\\
 6            & RLRLLC                      & 126.41-126.52 \\
 $\;\;$12     & RLRLLNRLRLLC                & 126.42 \\
 $\;\;$24     & RLRLLNRLRLLRRLRLLNRLRLLC    & 126.41 \\
 12           & RLRLRDLRLRLC                & 123.56-123.63 \\
 8            & RLRLRLLC                    & 121.687-121.689 \\
\hline\hline
\end{tabular}
\end{scriptsize}
\end{center}
\end{table}

\addtocounter{table}{-1}
\begin{table}
\caption{Continued.}
\begin{center}
\begin{scriptsize}
\begin{tabular}{llc}
\hline\hline
 7            & RLRLRLC                     & 118.128-118.134 \\
 14           & RLRLRRDLRLRLLC              & 116.91-116.925 \\
 5            & RLRLC                       & 113.9-114.01 \\
 $\;\;$10     & RLRLNRLRLC                  & 113.9 \\
 10           & RLRRDLRLLC                  & 110.57-110.70 \\
 9            & RLRRLLRLC                   & 108.9778 \\
 7            & RLRRLLC                     & 107.618-107.625 \\
 14           & RLRRLRDLRLLRLC              & 106.746-106.757\\
  8           & RLRRLRLC                    & 104.185 \\
 16           & RLRRLRRDLRLLRLLC            & 103.632-103.636\\
 3            & RLC                         & 99.79-100.795 \\
  $\;\;\;$6   & RLNRLC                      & 99.629-99.78 \\
$\;\;$12      & RLNRLRRLNRLC                & 99.57 \\
  9           & RLNRLRRLC                   & 99.275-99.285 \\
 12           & RRMLRDLLNRLC                & 94.542-94.554 \\
 6            & RRDLLC                      & 92.51-93.20  \\
 6            & RRLLLC                      & 92.155-92.5  \\
$\;\;$12      & RRLLLNRRLLLC                & 92.066-92.154 \\
 12           & RRLLRDLLRRLC                & 90.163-90.20 \\
 8            & RRLLRLLC                    & 88.368 \\
  7           & RRLLRLC                     & 86.402 \\
 14           & RRLLRRDLLRRLLC              & 85.986-85.987 \\
  8           & RRLLRRLC                    & 84.3365 \\
  5           & RRLLC                       & 83.36-83.39 \\
 $\;\;$10     & RRLLNRRLLC                  & 83.35 \\
 10           & RRLRDLLRLC                  & 82.040-82.095 \\
  8           & RRLRLLLC                    & 81.317 \\
  6           & RRLRLC                      & 76.818-76.822 \\
 $\;\;$12     & RRLRLNRRLRLC                & 76.815 \\ 
 12           & RRLRRDLLRLLC                & 76.310-76.713 \\
  8           & RRLRRLLC                    & 75.1405 \\
  7           & RRLRRLC                     & 73.712 \\
 14           & RRLRRRDLLRLLLC              & 73.457 \\
  4           & RRLC                        & 71.41-71.52 \\
  $\;\;\;$8   & RRLNRRLC                    & 71.43 \\
  8           & RRRDLLLC                    & 69.724-69.839 \\
  8           & RRRLLRLC                    & 66.2046 \\
  9           & RRRLLRRLC                   & 65.5025 \\
  6           & RRRLLC                      & 64.895-64.898  \\
 $\;\;$12     & RRRLLNRRRLLC                & 64.8946 \\
 $\;\;$24     & RRRLLNRRRLLRRRRLLNRRRLLC    & 64.893 \\
 12           & RRRLRDLLLRLC                & 64.572-64.574 \\
  7           & RRRLRLC                     & 62.069 \\
 14           & RRRLRRDLLLRLLC              & 61.928 \\
  9           & RRRLRRLLC                   & 61.31497 \\
  8           & RRRLRRLC                    & 60.654 \\
  5           & RRRLC                       & 59.242-59.255 \\
 $\;\;$10     & RRRLNRRRLC                  & 59.24 \\
 10           & RRRRDLLLLC                  & 58.700-58.715 \\
  9           & RRRRLLLLC                   & 58.0763 \\
  9           & RRRRLLRLC                   & 56.53315 \\
  7           & RRRRLLC                     & 55.787 \\
 14           & RRRRLRDLLLLRLC              & 55.675 \\
  6           & RRRRLC                      & 52.455-52.459 \\
 $\;\;$12     & RRRRLNRRRRLC                & 52.455 \\
 12           & RRRRRDLLLLLC                & 52.245-52.248 \\
  8           & RRRRRLLC                    & 50.3038-50.3240 \\
  7           & RRRRRLC                     & 48.1181-48.1194 \\
 $\;\;$14     & RRRRRLNRRRRRLC              & 48.1187 \\
 14           & RRRRRRDLLLLLLC              & 48.027 \\
\hline\hline
\end{tabular}
\end{scriptsize}
\end{center}
\end{table}

\begin{table}
\caption{Some stable periodic orbits in the Lorenz equations associated
with secondary sheets of the dynamical foliations.}
\label{table5}
\begin{center}
\begin{scriptsize}
\begin{tabular}{llc}
\hline\hline
\multicolumn{1}{c}{Period} & \multicolumn{1}{c}{Sequence} & $r$ \\
\hline
   3          & RMN                         & 328.0838  \\
   3          & RMC                         & 327.58-327.88 \\
$\;\;\;$6     & RMRRMC                      & 327.3-327.5 \\
$\;\;$12      & RMRRMNRMRRMC                & 327.26  \\
$\;\;$24      & RMRRMNRMRRMRRMRRMNRMRRMC    & 327.2  \\
 10           & RMRRDLNLLC                  & 191.982-191.985 \\
$\;\;$20      & RMRRMLNLLRRMRRMLNLLC        & 191.9795 \\
  6           & RLRRMC                      & 183.0435  \\
$\;\;$12      & RLRRMRRLRRMC                & 183.0434 \\
$\;\;$24      & RLRRMRRLRRMNRLRRMRRLRRMC    & 183.04338 \\ 
  6           & RLNRMC                      & 168.2492 \\
$\;\;$12      & RLNRMNRLNRMC                & 168.249189 \\
  4           & RRMR                        & 162.1381 \\
$\;\;\;$8     & RRMRRRMC                    & 162.13806 \\
$\;\;$16      & RRMRRRMNRRMRRRMC            & 162.13804 \\ 
 6            & RRMRMC                      & 157.671066 \\
$\;\;$12      & RRMRMNRRMRMC                & 157.6710656 \\ 
$\;\;$24      & RRMRMNRRMRMRRRMRMNRRMRMC    & 157.6710654 \\
 6            & RRLRMC                      & 139.9238433 \\
 $\;\;$12     & RRLRMRRRLRMC                & 139.9238430 \\
$\;\;$24      & RRLRMRRRLRMNRRLRMRRRLRMC    & 139.9238428 \\ 
\hline\hline
\end{tabular}
\end{scriptsize}
\end{center}
\end{table}
\end{document}